\documentclass[twocolumn]{aastex631}

\usepackage[T1]{fontenc}
\usepackage{ae,aecompl}

\usepackage{graphicx}	
\usepackage{amsmath}	
\usepackage{amssymb}	
\usepackage{float}
\usepackage{wasysym}
\usepackage{hyperref}

\shorttitle{CARB Progenitors}
\shortauthors{Van \& Ivanova}

\graphicspath{{./}{figures/}}

\begin{document}

\title{Constraining Progenitors of Observed LMXBs Using CARB Magnetic Braking}

\correspondingauthor{Kenny X. Van}
\email{kvan@ualberta.ca}

\author[0000-0003-3862-5826]{Kenny X. Van}
\affiliation{Department of Physics, University of Alberta, Edmonton, AB, T6G 2E7, Canada}

\author[0000-0001-6251-5315]{Natalia Ivanova}
\affiliation{Department of Physics, University of Alberta, Edmonton, AB, T6G 2E7, Canada}




\begin{abstract}
We present a new method for constraining the mass transfer evolution of low mass X-ray binaries (LMXBs) - a reverse population synthesis technique. This is done using the detailed 1D stellar evolution code \texttt{MESA} (Modules for Experiments in Stellar Astrophysics) to evolve a high-resolution grid of binary systems spanning a comprehensive range of initial donor masses and orbital periods. We use the recently developed Convection And Rotation Boosted (CARB) magnetic braking scheme. The CARB magnetic braking scheme is the only magnetic braking prescription capable of reproducing an entire sample of well studied persistent LMXBs -- those with their mass ratios, periods and mass transfer rates that have been observationally determined. Using the reverse population synthesis technique, where we follow any simulated system that successfully reproduces an observed LMXB backwards, we have constrained possible progenitors for each observed well-studied persistent LMXB. We also determined that the minimum number of LMXB formations in the Milky Way is 1500 per Gyr if we exclude Cyg X-2. For Cyg X-2, the most likely formation rate is 9000 LMXB per Gyr. The technique we describe can be applied to any observed LMXB with well-constrained mass ratios, period and mass transfer rate. With the upcoming GAIA DR3 containing information on binary systems, this technique can be applied to the data release to search for progenitors of observed persistent LMXBs.
\end{abstract}




\section{Introduction}
\label{Introduction}

Low mass X-ray binaries (LMXBs) are binary systems that consist of a star that is overfilling its Roche Lobe -- the donor, and of a compact companion, a neutron star (NS) or a black hole (BH) -- the accretor. NS LMXBs are among the most studied binary systems in astronomy, with large catalogues dedicated to their observations \citep[e.g.,][]{Ritter2003, Liu2007}. The data presented in these catalogues can allow for the derivation of quantities such as the mass transfer (MT) rates, mass ratios of the two stars, binary orbital periods and, in some cases, the donors' effective temperatures inferred from observations. These properties are essential for theoretical studies of the formation and evolution of LMXBs and are regularly used to compare theoretically obtained systems to the observed ones \citep{Bhattacharya1991, Podsiadlowski2002, Van2019a}. 

There are two very different approaches to gaining insight
in the theoretical studies of LMXBs. One approach is to study a specific observed system \citep{Eggleton1983, Justham2006, Pavlovskii2016, Podsiadlowski2000, Rappaport1995, Verbunt1981}. The other approach is to model a larger population of binaries simultaneously, using assumptions for the initial properties of the initial systems, and their evolution. This approach is known as population synthesis and is commonly used to analyze the statistical properties of a type of system \citep{Breivik2020, Fragos2008, Kalogera1998, Kobulnicky2007, Rappaport2005}. Population synthesis codes have been used to calculate formation rates for a wide range of exotic stellar systems using different initial conditions \citep{Belczynski2018, Belczynski2020, Bruzual2003}. In standard population synthesis studies, initial conditions such as the initial mass distribution and the stellar birth rates control the formation and evolution of stellar systems. Depending on the type of systems studied, various other parameters may be adjusted, such as the wind mass-loss rates, accretion luminosity equations, and common envelope prescriptions. The validity of population synthesis is in the accuracy of these initial conditions and parameters or prescriptions used.

In this work, we will be employing a {\it reverse} population synthesis technique. Instead of applying a set of initial conditions to analyze the obtained systems during and after their evolution, we will use our simulated results to infer the initial progenitor properties.

The theoretically obtained LMXBs will be used to constrain the possible progenitor conditions of our observed systems and estimate the required formation rate of a given progenitor. In Section \S \ref{sec:theory} we review the simulation setup -- the grid of binaries we used and conditions used to evolve the systems. Section \S \ref{sec:analysis} presents the viable progenitors of each subgroup of LMXBs. Section \S \ref{sec:rates} presents the estimates for the formation rate of the progenitor binaries. In Section \S \ref{sec:unobserved} we take a closer look at the gaps in our parameter space that do not result in any observed LMXBs and discuss why these binaries aren't seen, and we provide the predictions on where future LMXBs may be detected.

\section{Model}
\label{sec:theory}

The detailed numerical setup of how we model a population of LMXBs was described in \cite{Van2019a}. Below we provide the most crucial points or changes relevant to the presented study.

\subsection{Population Grid}
\label{subsec:numerical_setup}

Following \cite{Van2019a} and \cite{Van2019b}, we create an initial grid of NS LMXBs with a range of initial periods and masses while using a higher-resolution grid as compared to the previous studies. The parameters of the mesh are:

\begin{itemize}
    \item Initial donor masses are in the range $0.95 \leq M_{\rm d}/M_\odot \leq 7.00$.
    \item Initial binary period are in the range  $-0.6 \leq \log_{10}(P/\rm day) \leq 4.0$.
\end{itemize}

We use a non-uniform mesh density for the initial donor mass and period, see Table \ref{tab:mesh} for the distribution. With the adopted mesh density, we model a total of 14836 binary systems.

All donors are placed in binary systems with a NS accretor that has an initial mass of $M_{\rm a} = 1.4 M_\odot$ and radius $R_{\rm a} = 11.5\rm\ km$. A binary system at any point in its evolution can also be characterized by the derived quantity $q = M_d/M_a$, which is the mass ratio between the donor and accretor. All donors are initially at their zero-age main sequence. For all simulations, we adopt the default \texttt{MESA} metallicity of $Z = 0.02$, while solar metallicity has been suggested to be lower at $Z_\odot \approx 0.13$ \citep{Asplund2009}\footnote{We have verified that reducing the metallicity by a factor of two does not significantly affect the results presented elsewhere in this paper.}. The chosen grid of systems covers all binaries that can start mass transfer at some point during their evolution, with lower mass donors unlikely to contribute to the population of LMXBs during a Hubble time.

\begin{table}
    \centering
    \begin{tabular}{c c c c}
        $M_{\rm d}$ range         & $ \Delta M_{\rm d} /M_\odot$ & $\log_{10}(P)$ range     & $\Delta \log_{10}(P/{\rm day})$ \\
        \hline
        $$ 0.95 -- 4.00 $$ & $ 0.05 $     & $$ -0.6 -- 1.64 $$ & $ 0.02 $ \\
        $$ 0.95 -- 4.00 $$ & $ 0.05 $     & $$ 1.65 -- 4.0 $$  & $ 0.05 $ \\
        $$ 4.00 -- 7.00 $$ & $ 0.10 $     & $$ -0.6 -- 1.64 $$ & $ 0.02 $ \\
        $$ 4.00 -- 7.00 $$ & $ 0.10 $     & $$ 1.65 -- 4.0 $$  & $ 0.05 $ \\
    \end{tabular}
    \caption{The grid sizes for mass and orbital periods  for different ranges of initial donor masses and initial orbital periods. 
    }
    \label{tab:mesh}
\end{table}

The simulations were performed using the one-dimensional stellar evolution code \texttt{MESA}\footnote{\url{http://mesa.sourceforge.net}} (Modules for Experiments in Stellar Astrophysics) revision 11701 \citep{Paxton2011, Paxton2013, Paxton2015, Paxton2018, Paxton2019}, and May 2019 release of  \texttt{MESASDK} \citep{Townsend2019}. The wind mass loss scheme used in our donor star is the ``Reimers'' wind mass prescription \citep{Reimer1975} with no additional boosting factors included. We also allow for the evolution of the radial velocity in our donor star. 

For our binary parameters, we use a modified Eddington limit appropriate for NSs and an improved magnetic braking prescription. Both the magnetic braking prescription and the modified Eddington limit will be described in further detail in Sections \ref{subsec:MB_derivation} and \ref{subs:accr} respectively. Beyond these changes in our simulation parameters, the other properties are done using \texttt{MESA} defaults.

The MESA EOS is a blend of the OPAL \citep{Rogers2002}, SCVH \citep{Saumon1995}, PTEH \citep{Pols1995}, HELM \citep{Timmes2000}, and PC \citep{Potekhin2010} EOSs.

Radiative opacities are primarily from OPAL \citep{Iglesias1993, Iglesias1996}, with low-temperature data from \citet{Ferguson2005} and the high-temperature, Compton-scattering dominated regime by \citet{Buchler1976}.  Electron conduction opacities are from \citet{Cassisi2007}.

Nuclear reaction rates are a combination of rates from NACRE \citep{Angulo1999}, JINA REACLIB \citep{Cyburt2010}, plus additional tabulated weak reaction rates \citep{Fuller1985, Oda1994, Langanke2000}. Screening is included via the prescriptions of \citet{Salpeter1954, Dewitt1973, Alastuey1978} and \citet{Itoh1979}. Thermal neutrino loss rates are from \citet{Itoh1996}.

The orbital evolution of the binary systems is governed by the total angular momentum loss through three main mechanisms. These mechanisms include angular momentum loss due to magnetic braking (see \S\ref{subsec:MB_derivation}), through angular momentum loss via mass loss from the system (see \S\ref{subs:accr}), and in very short period systems, angular momentum loss due to gravitational radiation \citep{Faulkner1971}. Our simulations do not account for any additional effects the compact object may have on the donor star, such as irradiation or tidal heating.

When the donor star overfills its Roche lobe \citep[for a one-dimensional approximation of the Roche lobe radius, see ][]{Eggleton1983}, the material will flow from the donor to the accretor through the $L_1$ Lagrange point between the two stars. The rate of mass transfer (MT) is calculated using the  ``Ritter" prescription in \texttt{MESA} \citep[for more information on the MT prescription, see][]{Ritter1988}. The details on the accretion rate and the Eddington limit will be described in section \S \ref{subs:accr}.

The simulated LMXBs all start with an NS already formed and the donor star at zero-age main sequence (ZAMS). All simulations continued for 10 Gyrs, or until the donor star detaches and no longer transfers mass to the NS. In some cases, the simulation would experience dynamically unstable mass transfer which \texttt{MESA} is not designed to simulate. These simulations would likely encounter numerical issues and stop. Some of the systems were initially placed at such a short period that they overfill, sometimes significantly, their Roche lobe at the start of simulations while surviving this initial MT. We did not discard these systems but discuss them separately with caution.

\subsection{Magnetic Braking}
\label{subsec:MB_derivation}

We use the Convection and Rotation Boosted (CARB) magnetic braking \citep[see][for the derivation]{Van2019b}. The angular momentum loss from a star with radius $R$, rotating with a rate of $\omega$ at its surface,  is:

\begin{equation}
\begin{split}
    \Dot{J}_{\rm MB}=&-\frac{2}{3}\dot{M}_{\rm W}^{-1/3} R^{14/3} \left( v_{\rm esc}^2 + \frac{2 \Omega^2 R^2}{K_2^2}\right )^{-2/3}\\
    &\times\Omega_\odot\  B_{\odot}^{8/3}\ \left(\frac{\Omega }{\Omega_\odot}\right)^{11/3}\left(\frac{\tau_{\rm conv} }{\tau_{\odot, \rm conv}}\right)^{8/3}\ .
\label{eq:CARB_MB}
\end{split}
\end{equation}

\noindent $\Dot{M}_{\rm W}$ denotes the wind mass loss rate, $v_{\rm esc}$ is the surface escape velocity, $\tau_{\rm conv}$ is the convective turnover time, $B$ is the surface magnetic field strength, and $K_2=0.07$ in this equation is a constant obtained from a grid of simulations by \cite{Reville2015}. $K_2$ sets the limit to the rotation rate required to play a significant role in damping the magnetic braking. The solar values adopted to normalize Equation \ref{eq:CARB_MB} are $B_\odot = 1\rm\ G$ for Sun's surface magnetic field strength, and $\Omega_\odot \approx 3\times10^{-6} \ \rm s^{-1}$ for Sun's surface rotation rate. For convective turnover time we use $\tau_{\odot, \rm conv}=2.8 \times 10^6 \ \rm s$. This value was obtained by evolving a $1M_\odot$, $Z=Z_\odot$ star to $4.6 \rm\ Gyrs$ and using the following equation:

\begin{equation}
\tau_{\rm conv} = \int_{R}^{R_{s}} \frac{dr}{v_{\rm conv}} .
\end{equation}

\noindent $R$ and $R_s$ are the bottom and the top of the outer convective zone respectively, while $v_{\rm conv}$ is the local convective velocity \citep[for more details, see][]{Van2019a}.

\subsection{Accretion rate}
\label{subs:accr}

Following \cite{Van2019a} and \cite{Van2019b}, we consider non-conservative MT, where we limit the mass accretion rate by the Eddington-limited maximum accretion rate. If the MT rate exceeds the Eddington limit, the excess is not considered to be accreted by the compact object but is counted as lost from the system with the accretor's specific angular momentum. The Eddington-limited mass accretion rate $\dot{M}_{\rm Edd}$ is:

\begin{equation}
    \dot{M}_{\rm Edd }=\frac{4\pi c R_{\rm a}}{\kappa_{e}} \approx \frac{3.4}{1+X}\times 10^{-8} M_\odot \rm {\ yr}^{-1} 
    \label{eq:MT_Edd}
\end{equation}

\noindent Here, $\kappa_e$ is the Thomson electron scattering opacity, $\kappa_{e}=0.19(X+1)$ cm$^2$ g$^{-1}$, $X$ is the hydrogen mass fraction in the material transferred from the donor.

In addition to the Eddington limit, we also impose MT efficiency $\eta$ which represents a mass transfer efficiency. An analytic description of $\eta$ is not currently known, but previous work has shown that this efficiency can range between $\sim 0.05 - 0.3$ in pulsars \citep{Antoniadis2012}. Previously, \cite{Van2019b} showed that an efficiency $\eta = 1$ would result in neutron star masses consistently exceeding $2M_\odot$ and an efficiency of $\eta = 0.2 $ results in a more reasonable upper mass value of $\sim 1.8 M_\odot$ \citep[for an explanation see][]{Van2019b}. Combining the Eddington limit and the mass transfer efficiency, the material accreted by the NS is

\begin{equation}
\begin{split}
    \dot{M}_{\rm NS} = \text{min}(\dot{M}_{\rm Edd},\ \eta \dot{M}_{\rm tr})
\end{split}
\end{equation}

\noindent $\dot{M}_{\rm tr}$ is the mass lost by the donor via its $L_1$ Lagrange point due to Roche lobe overflow. The amount of material accreted onto the NS determines the observed luminosity of the system. We do not set an upper limit on the mass loss rate of the donor and only limit the accretion rate of the NS. It is important to note material accreted onto the disc that moves closer to the compact object but is not accreted onto the compact object itself will also contribute to the luminosity of the system. Along with MT efficiency $\eta$ not being constant throughout the evolution, we estimate that the luminosity of our systems can be described by the equation: 

\begin{equation}
    L = \frac{0.6 G \dot{M_{\rm tr}} M_{\rm NS}}{R_{\rm NS}}
\end{equation}

\noindent To account for these uncertainties, the sizes of our MT bins are large enough to compensate during analysis.

\subsection{Persistent Systems}
\label{subsec:DIM}

To compare the simulated systems to the observed systems, we must determine if the simulated system is observable. Specifically, we only compare to the systems which are observationally classified as persistent LMXBs. To discriminate whether the modelled system would be deemed persistent or transient from observations,  we use the disc instability model (DIM). The DIM states that there exists a critical mass transfer rate that separates persistent and transient systems at a given period \citep{Meyer1981}. In systems where the accretion is too low, the created accretion disc experiences a buildup of material. The system will then experience outbursts separated by the periods of quiescence caused by increases and decreases in temperature, and thus appears as a transient system. Conversely, if the accretion rate is high for this orbital period, the accretion disc is in a constantly bright state, appearing persistent \citep[see][for a review of DIM]{Lasota2001}. In this paper, we use the stability criteria from \cite{Coriat2012}:

\begin{equation}
\begin{split}
    \dot{M}_{\rm crit} = k P^b_{\rm hr}  
\label{eq:DIM}
\end{split}
\end{equation}

\noindent $P_{\rm hr}$ is the orbital period of the binary system in hours. 
The most lenient classification that predicts the largest number of persistent systems requires an irradiated disc with a neutron star accretor. The condition for an irradiated disc instability results in the lowest $\dot{M}_{\rm crit}$ for any given orbital period. The condition of instability for an irradiated accretion disc is described by Equation~\ref{eq:DIM} with $b = 1.59$ and $k = (2.9 \pm 0.9) \times 10^{15}  \mathrm{g\ s}^{-1}$. In our calculations we will be using the lower limit of $k = 2.0 \times 10^{15} \mathrm{g\ s}^{-1}$ from \cite{Coriat2012} in determining the critical mass transfer rate to capture the largest number of simulated persistent systems.  Any MT system that exceeds this critical transfer rate is classified in our simulations as a persistent LMXB. This MT rate is the amount of mass flowing through the $L_1$ point.

\subsection{Relating the observed LMXBs to the simulated LMXBs}

For our analysis, we split the sample of observed binaries based on the orbital period (see Table ~\ref{table:combined_table}). Ultra-compact X-ray binaries (UCXBs) are LMXBs defined to have an orbital period of shorter than 80 minutes. Systems with periods less than 4 hours are often classified as short period LMXBs. Our sample only contains one binary with a period of about ten days, Cyg X-2, which we define as a long period binary. We define LMXBs with orbital periods ranging from 10 hour to approximately 1 day as {\it medium period} systems for the analysis in this paper.

To determine the progenitors of the observed LMXBs of interest, we must simultaneously match the orbital period, the MT rate, and the mass ratio of our simulated system with one of the observed systems. We treat each observed, persistent LMXB not as occupying a single point in this three-dimensional space but as occupying a cuboid in this space. Each cuboid is centred at one of the observed LMXBs. The lengths of the cuboid edges are taken to be equal to the uncertainties in each of the observed quantities for this observed LMXB. In the case of the mass ratio and mass transfer rates, these are the observed errors in each quantity while the length of $\log_{10}(P)$ represents a reasonable range for the value. See Table~\ref{table:combined_table} for details on the adopted cuboids in the period, mass ratio, and MT rate space. These centres of each cuboid and the errors in each dimension are taken from \cite{Van2019b}. 

For LMXBs with additional observed properties, we can further constrain the number of possible progenitors. In the case of Sco X-1 and Cyg X-2, there are observation values for the effective temperatures of the donor star. Observations by \cite{Sanchez2015} found that the donor star in Sco X-1 was later than a K4. This observation constrains the donor star's effective temperature to have an upper limit of $T_{\rm eff} \lesssim 4800 \rm\ K$. Cyg X-2, on the other hand,  is better constrained as the observations by \cite{Cowley1979}, which limit the effective temperature to $7000 \lesssim T_{\rm eff}/\rm K \lesssim 8500$. With these additional observed quantities, we can compare the effective temperature of our simulated systems to these observed constraints to limit the viable progenitors.
 
\begin{table*}
\footnotesize
\caption{\textbf{Binned Properties of LMXBs}}
\centering
\begin{tabular}{l  lllll}

System Name          & $\log_{10}(P/\rm day)$ & $q$            & $\log_{10}(\dot M_{\text{tr}})$ & $\tau_{\rm max}$ (years) & $A_{\rm sys}/A_{\rm tot}$ \\
\hline
\multicolumn{6}{l}{UCXB} \\
4U 0513-40           & [-1.95, -1.90]         & [0.01, 0.06] & [-9.0, -8.4]          & $ 1.08\times10^{7}     $ & $ 1.37\times 10^{-3}    $ \\ 
2S 0918-549          & [-1.94, -1.89]         & [0.01, 0.06] & [-9.6, -8.4]          & $ 1.12\times10^{7}     $ & $ 1.41\times 10^{-3}    $ \\ 
4U 1543-624          & [-1.92, -1.87]         & [0.01, 0.06] & [-8.9, -8.4]          & $ 1.44\times10^{7}     $ & $ 1.44\times 10^{-3}    $ \\ 
4U 1850-087          & [-1.86, -1.81]         & [0.01, 0.06] & [-9.8, -8.2]          & $ 2.34\times10^{7}     $ & $ 1.69\times 10^{-3}    $ \\ 
M15 X-2              & [-1.82, -1.77]         & [0.01, 0.06] & [-9.5, -8.9]          & $ 3.22\times10^{7}     $ & $ 1.76\times 10^{-3}    $ \\ 
4U 1626-67           & [-1.55, -1.50]         & [0.01, 0.06] & [-9.5, -8.4]          & $ 9.33\times10^{7}     $ & $ 2.25\times 10^{-3}    $ \\ 
4U 1916-053          & [-1.48, -1.43]         & [0.03, 0.08] & [-9.4, -8.7]          & $ 5.82\times10^{7}     $ & $ 7.40\times 10^{-4}    $ \\ 
\\
\multicolumn{6}{l}{Short period}  \\
4U 1636-536          & [-0.82, -0.77]         & [0.15, 0.40] & [-8.9, -8.4]          & $ 2.53\times10^{7}     $ & $ 4.05\times 10^{-3}    $ \\  
GX 9+9               & [-0.78, -0.73]         & [0.20, 0.33] & [-8.5, -8.0]          & $ 1.66\times10^{7}     $ & $ 3.91\times 10^{-3}    $ \\  
4U 1735-444          & [-0.73, -0.68]         & [0.29, 0.48] & [-8.2, -7.7]          & $ 1.07\times10^{7}     $ & $ 4.23\times 10^{-3}    $ \\  
2A 1822-371          & [-0.65, -0.60]         & [0.26, 0.36] & [-7.6, -7.1]          & $ 6.15\times10^{6}     $ & $ 5.32\times 10^{-3}    $ \\  
\\
\multicolumn{6}{l}{Medium period}  \\
Sco X-1              & [-0.12, -0.07]         & [0.15, 0.58] & [-7.8, -7.1]          & $ 9.42\times10^{6}     $ & $ 1.66\times 10^{-3}    $ \\  
GX 349+2             & [-0.05, 0.00]          & [0.39, 0.65] & [-7.8, -7.1]          & $ 1.58\times10^{7}     $ & $ 4.37\times 10^{-3}    $ \\  
\\
\multicolumn{6}{l}{Long period}  \\
Cyg X-2              & [0.97, 1.02]           & [0.25, 0.53] & [-7.8, -7.0]          & $ 8.36\times10^{5}     $ & $ 2.30\times 10^{-3}    $ \\  
\hline

\end{tabular}
\label{table:combined_table}
\begin{flushleft}
\textbf{Notes.} The binned properties of the observed LMXBs used to compare to simulated systems and diagnostic properties used to analyze the results. The binned properties are period in days, mass transfer rate in $\dot M_{\text{tr}}$ is in $M_\odot\ \mathrm{yr}^{-1}$ and the mass ratio. The bin ranges are centred on the observed values and span the errors in the given observed property. The two diagnostic properties are the maximum amount of time a given simulated system spends in the observed bin of interest, $\tau_{\rm max}$ and $A_{\rm sys}/A_{\rm tot}$ is the fraction of our tested parameter space that can reproduce the system of interest. These two quantities indicate how long a simulation appears similar to an observed LMXB and how many systems could reproduce these properties. The period, mass ratio and mass transfer rate bins are taken from Table 4 from \cite{Van2019a}.
\end{flushleft}
\end{table*}

\subsection{Progenitor search}

To analyze the progenitor population, we start with finding the total observed time  $\tau_{\rm observed}$, which is defined as the total amount of time that a seed progenitor binary spends while appearing similar to any of the observed binaries, e.g., it passes through any of the cuboids. 
We then find the amount of time a simulated binary spends as a persistent binary predicted by Equation \ref{eq:DIM}, $\tau_{\rm persistent}$. Their ratio is:

\begin{equation}
    f_{\rm obs} = \frac{\text{ Time matching any observed system}}{\text{Time satisfying persistent condition}} =\frac{\tau_{\rm observed}}{\tau_{\rm persistent}},
    \label{eq:obs_frac}
\end{equation}

\noindent The amount of time our simulations spend as a persistent system exceeds the amount of time it appears similar to any of the observed binaries. This ratio gives us an idea of how likely it is that a progenitor will be able to produce an observed system. If a progenitor has a very high ratio, this means that the system spends a significant fraction of its persistent lifetime appearing similar to an observable LMXB. We will also define an additional parameter $f_{\rm{obs},i}$:

\begin{equation}
    f_{\text{obs},i} = \frac{\text{Time matching a specific observed system}}{\text{Time satisfying persistent condition}}
    \label{eq:obs_frac_i}
\end{equation}

\noindent Using this definition $f_{\text{obs}, i} \leq f_{\rm obs}$. We will use $f_{\text{obs},i}$ in figures where we will be looking at progenitors of a single observed LMXB and $f_{\rm obs}$ for figures where we determine the progenitor of multiple LMXBs.

\section{Analysis}
\label{sec:analysis}

\begin{figure}
    \centering
    \includegraphics[width=\columnwidth]{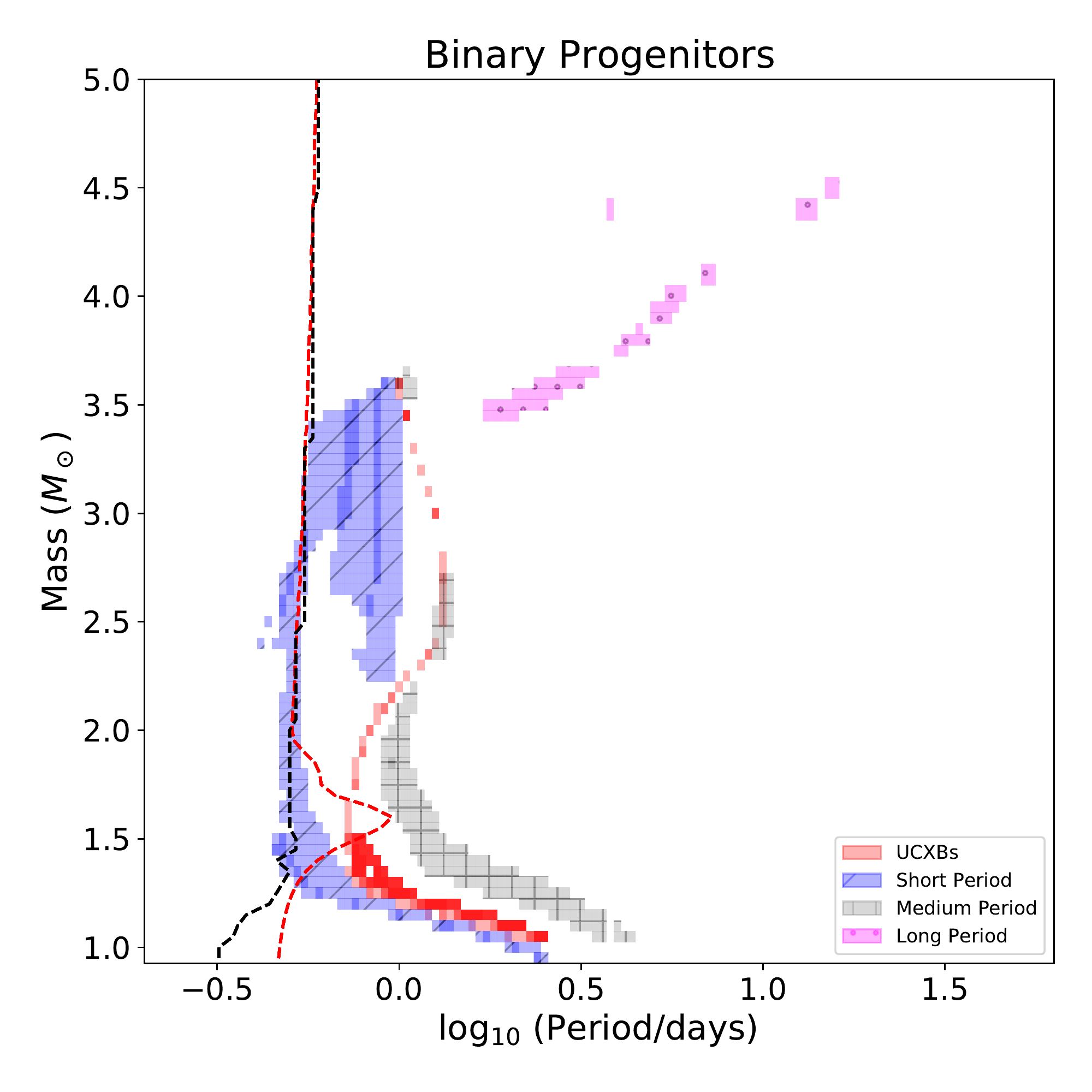}
    \caption{The progenitors of the observed LMXBs split into distinct classes based on the periods of the currently observed systems as denoted in Table \ref{table:combined_table}. Some observed systems share common progenitors; to show this, the individual grid points are semitransparent. The black dashed line denotes the shortest initial period with which the initial binary would start as a detached system. The red dashed line denotes the shortest initial period which the detached binary can have when the progenitor is 10 million years old. The systems on the left of the black dashed line started their evolution while having an initial Roche lobe overflow.}
    \label{fig:Progenitor_class}
\end{figure}

\begin{figure*}
    \centering
    \includegraphics[width=\textwidth]{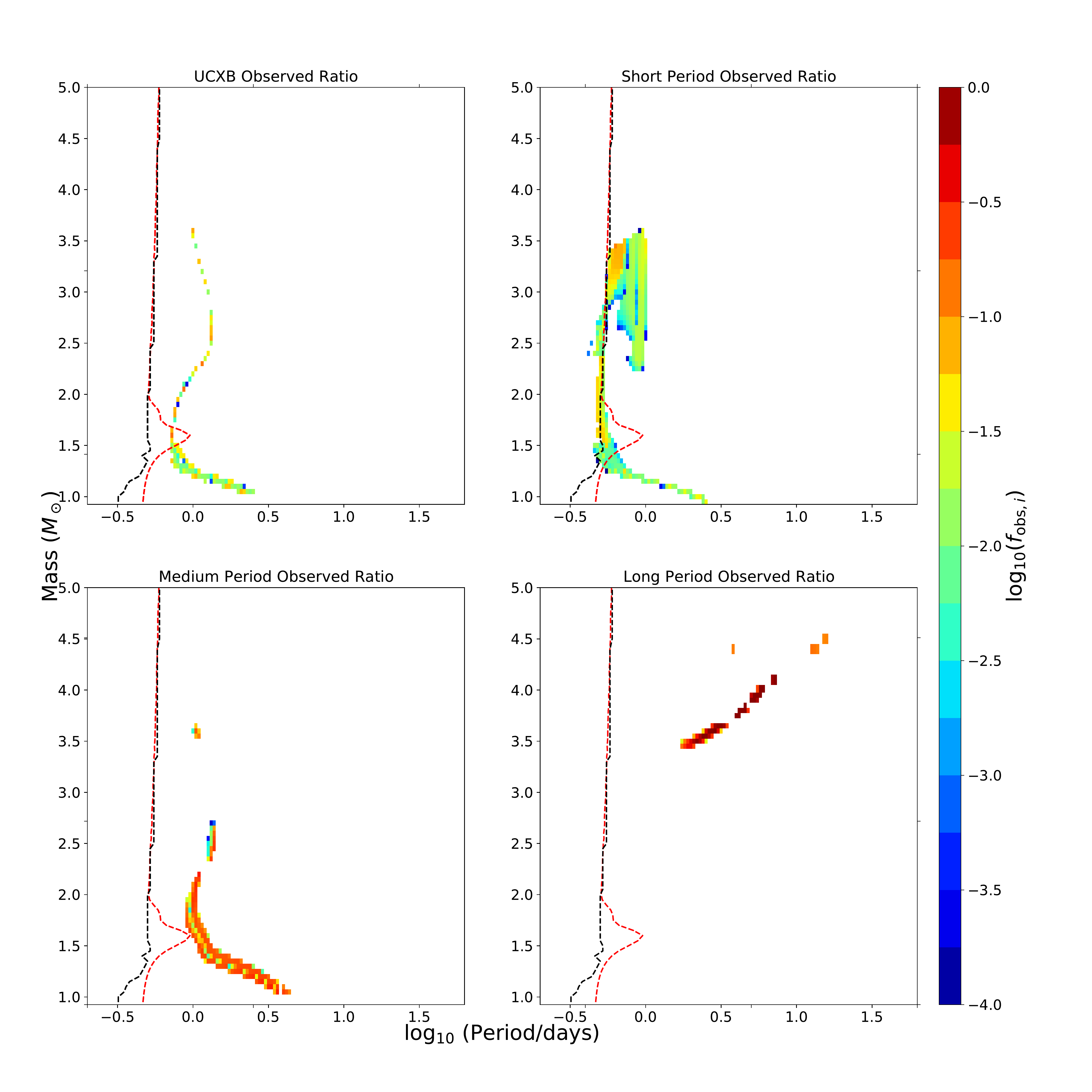}
    \caption{The progenitors of the four types of the observed LMXB. The colour shows $f_{\rm obs, i}$, the fraction of time the modelled system shows up as one of the observed LMXB, as compared to the time it can be observed as any persistent LMXB, see Equation \ref{eq:obs_frac}). The black and the red dashed lines are as in Figure \ref{fig:Progenitor_class}.}
    \label{fig:Progenitor_Grid}
\end{figure*}

\begin{figure*}
    \centering
    \includegraphics[width=0.49\textwidth]{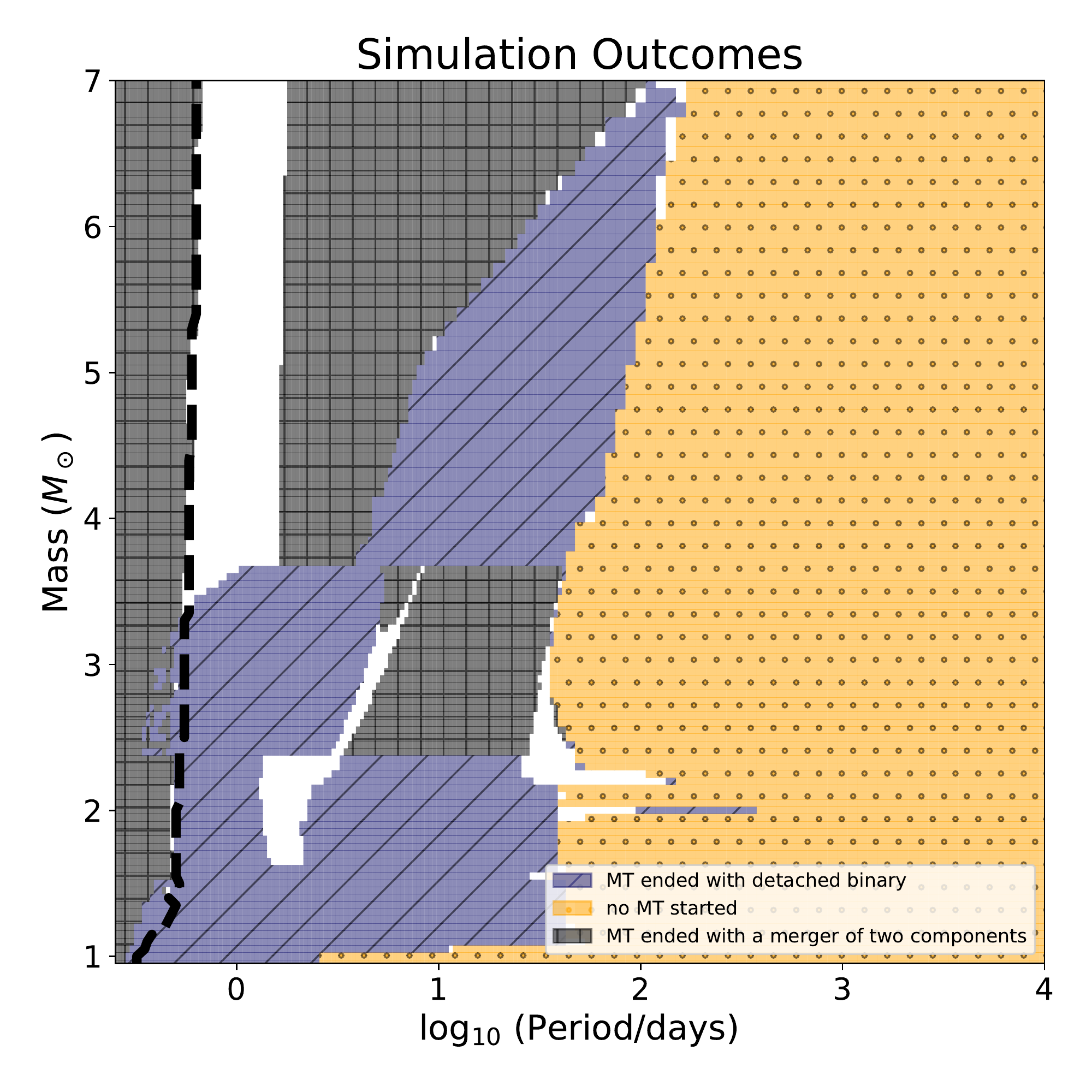} 
    \includegraphics[width=0.49\textwidth]{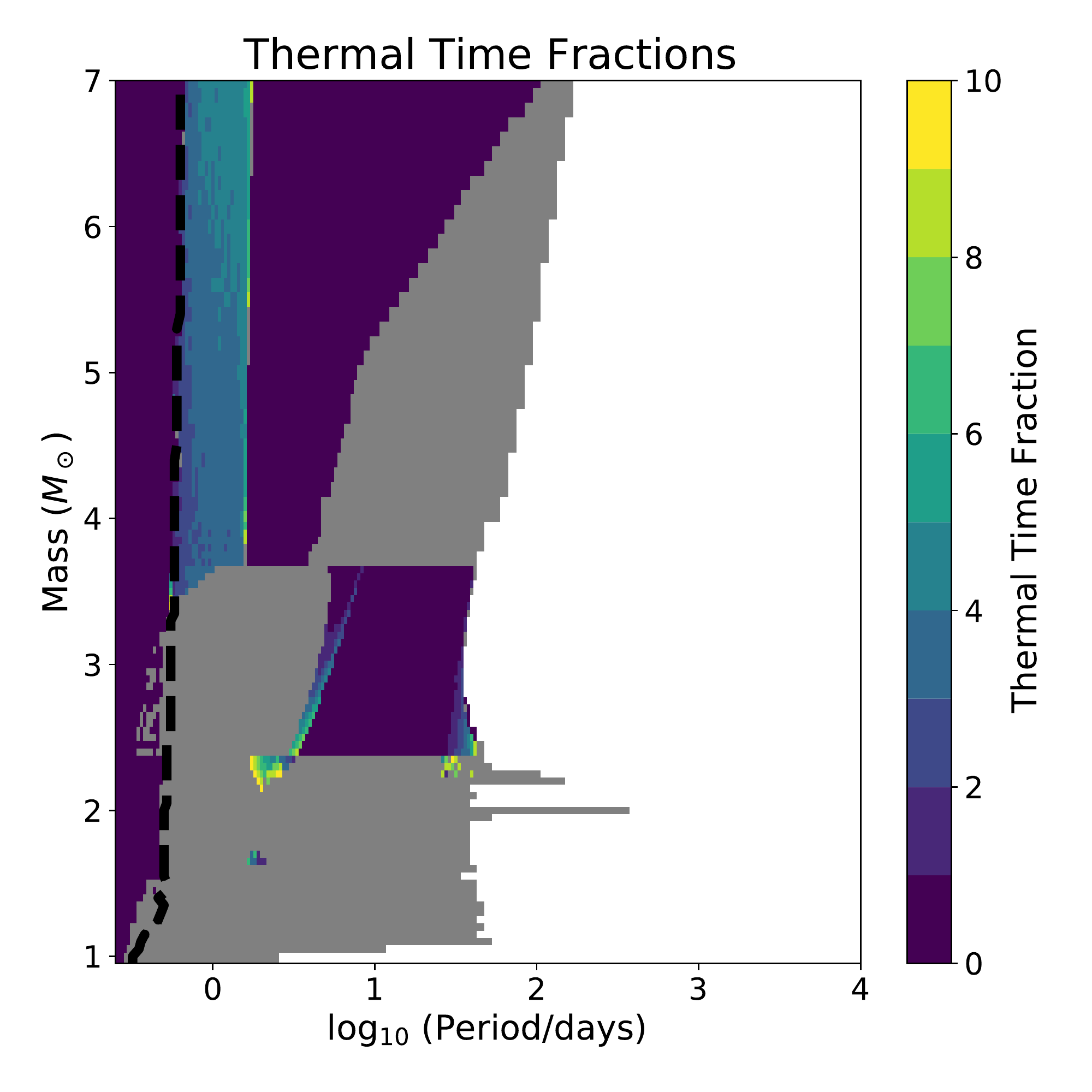} 
    \caption{The left figure shows the outcome of the entire simulation with a given progenitor donor mass and orbital period. The black dashed line denotes the largest initial orbital period where the simulated binary starts with an overfilled Roche lobe. The white regions denote regions where the simulated systems experience numerical issues with {\texttt MESA}, as they experience exponential growth in mass transfer exceeding the thermal mass transfer rate by at least two orders of magnitude. This takes place after a period of stable thermal timescale mass transfer, but the eventual outcome of these mass transferring systems -- a merger or a binary -- is not clear. The right figure shows the thermal timescale fractions for a simulated binary. The grey colour denotes where the thermal timescale fractions exceed 10, and the white denote regions where no mass transfer occurs.}
    \label{fig:sim_result}
\end{figure*}

We present the progenitors of the observed systems in Figure \ref{fig:Progenitor_class}. While the initial parameter space spans a significant range in mass and period, the viable progenitors are constrained to a small area of the parameter space. Specifically, only the binary systems with the initial donor masses between  $0.95M_\odot$ and $4.5M_\odot$ and with the orbital periods between 0.4 days and 16 days have been found to contribute to any of the existing persistent LMXBs with only a portion of the simulated systems initiating mass transfer and ending the simulation as a detached system as seen in Figure \ref{fig:sim_result}. The progenitors of the individual types of LMXBs form distinct groups. Below, we will examine each of them in detail.

We anticipate that in field binaries, some companions are still in the pre-main sequence stage when NS formation takes place$^\ddag$. For that reason, we also analyze the constraints on the progenitor space due to the minimum period which a binary could have when the progenitor-donor is 10 million years old (an NS could have been created between 7 and 25 million years after the two stars started their formation, depending on the initial mass of the initially more massive star). Specifically, at the age of 10 million years, only stars more massive than $2 M_\odot$ have reached the ZAMS (see Figure~ \ref{fig:Progenitor_class}). At the same time, pre-MS stars with about $1.6 M_\odot$ separate the stars with radiative and convective outer zones, resulting in a bump in the plausible initial periods.

\subsection{Ultra Compact Systems}

By splitting our sample of observed binaries into the period ranges given in Table \ref{table:combined_table}, we can determine if there are specific ranges in our parameter space that reproduce a subset of the observed systems. The progenitors which reproduce the observed UCXBs are shown in Figure \ref{fig:Progenitor_Grid}.

In LMXB evolution, for each donor and accretor combination, there is a critical value of the initial orbital period known as the bifurcation period. The bifurcation period separates the binaries where mass transfer leads to shrinking orbital separations and those where the orbital period increases, at least until the donor detaches and MT no longer occurs. The progenitors of UCXBs in the field accordingly are considered to have the initial periods below the bifurcation period for their companions \citep{Nelson1986, Nelson2003, Podsiadlowski2002}. The bifurcation periods' values are also a function of the accepted magnetic braking. Our results for UCXBs progenitors are in complete agreement with their standard formation scenario.

The entire group of UCXBs progenitors comes from a very narrow portion of the parameter space with very few viable initial periods for a given initial donor mass. The initial period of UCXBs has the largest possible range for donors with lower masses spanning up to $\log_{10}(P/\rm days) \approx 0.2$. As the mass increases, this period range decreases to $\log_{10}(P/\rm days) \approx 0.02$. From the maximum amount of time spent in an observed bin listed in Table \ref{table:combined_table}, we see that the modelled systems spend a significant amount of time appearing similar to the observed UXCBs. Having a large $f_{\rm obs}$ is expected, as these binaries are highly evolved systems near the end of the donor stars' life, and the evolution of UCXBs slows down almost exponentially with time.

The two UCXBs, 4U 1626-67 and 4U 1916-053 have longer detected periods than the other five ultra-compact systems. These two systems have been found to require more complex stellar evolution, such as enhanced angular momentum loss at shorter periods, an evolved main sequence donor with finely tuned initial parameters for either magnetic braking, or common envelope evolution \citep{Heinke2013, Podsiadlowski2002}. Our progenitors that successfully reproduce 4U 1626-67 and 4U 1916-053 are all main sequence stars that have not evolved enough to produce a significant amount of helium in their cores. Our formation channel is an alternative evolutionary channel, where the updated magnetic braking results in the donor star having sufficiently high angular momentum loss at periods less than 80 minutes and drive up the MT rate. In our simulations, reproducing these two UCXBs also requires fine-tuning in the sense that the initial progenitor space is very small, with a narrow region of possible donors more massive than $1.50 M_\odot$ contributing. The other UCXBs do not contain any progenitors with with initial masses exceeding $1.50 M_\odot$.

\subsection{Short Period}

\begin{figure*}
    \centering
    \includegraphics[width=\textwidth]{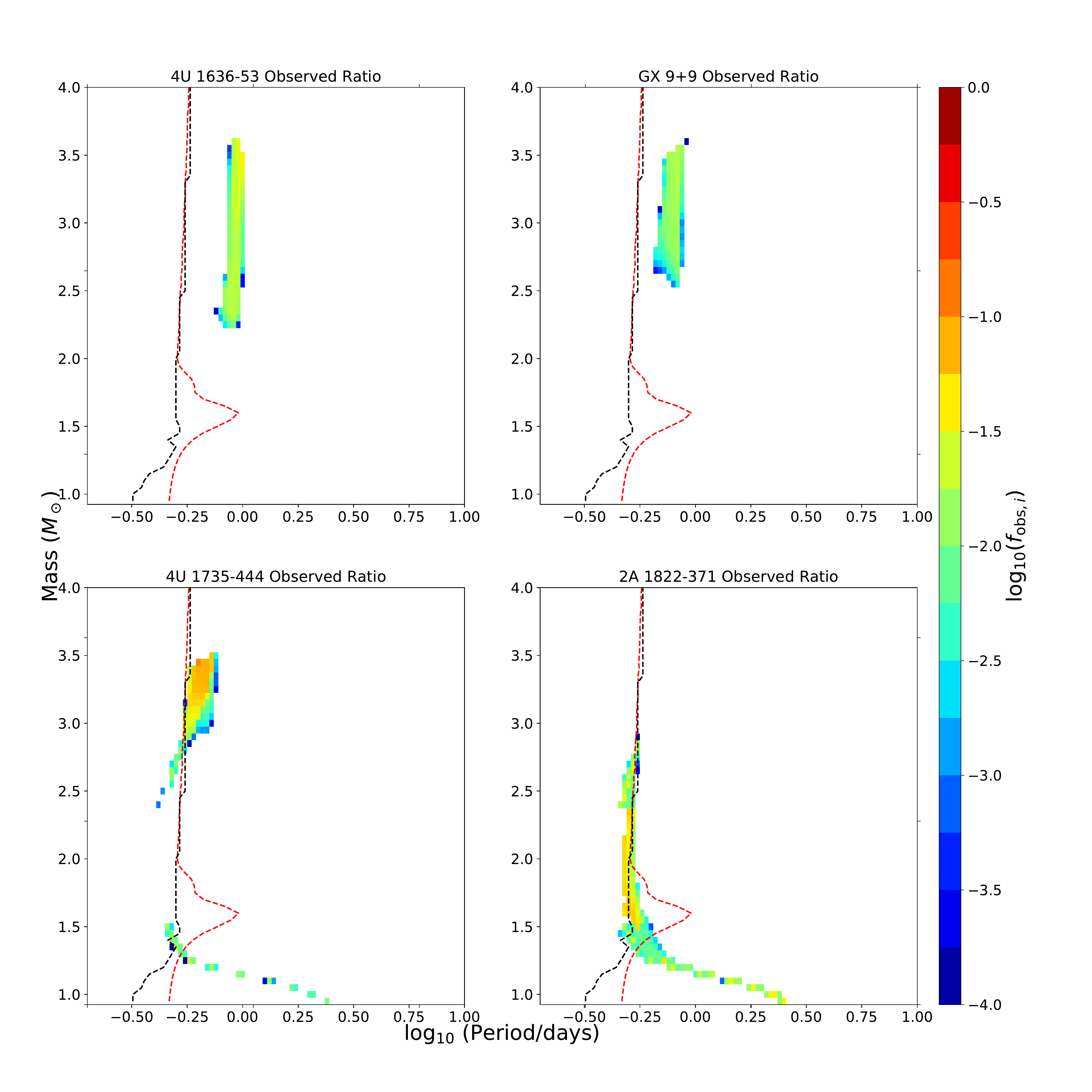}
    \caption{The short period systems from Figure \ref{fig:Progenitor_Grid} split into the individual systems. The colour shows $f_{\rm obs}$, the fraction of time the modelled system shows up as one of the observed LMXB, as compared to the time it can be observed as any persistent LMXB. The black and the red dashed lines are as in Figure \ref{fig:Progenitor_class}.}
    \label{fig:short_period_split}
\end{figure*}

The progenitors of the short period systems span a much larger period and mass range than the ultra-compact systems, with a larger fraction of the progenitor space resulting in an observed binary (see Figure~\ref{fig:Progenitor_Grid}). Initial donor masses that form our short period systems range from $0.9 M_\odot$ to  $3.6 M_\odot$, with the periods between 0.45 day and 2.5 days ($-0.34 \leq \log_{10}(P/ \rm days) \leq 0.40$). Unlike the UCXB systems, there is no clear pattern or structure in the initial parameter space needed to reproduce all of the observed short LMXBs. Plotting the progenitors for each observed short LMXB system individually demonstrates their distinct differences (see Figure~\ref{fig:short_period_split}).

The progenitors of 4U~1636-536 and GX~9+9 appear to belong to very confined progenitor groups.  The masses are confined to $2.25 \leq M/M_\odot \leq 3.60$ and $2.55 \leq M/M_\odot \leq 3.60$, respectively, and the initial periods spans are $-0.12 \leq \log_{10}(P/ \rm days) \leq 0$ and $-0.18 \leq \log_{10}(P/ \rm days) \leq -0.04$, respectively. These two LMXBs, 4U 1636-53 and GX 9+9, share progenitors with initial masses ranging from $2.70 \leq M/M_\odot \leq 3.55$ and $\log_{10}(P/ \rm days) = -0.06$.

The progenitors of 4U~1735-444 are split into two distinct groups based on the mass, with the higher mass group ranging from $2.50 \lesssim M/M_\odot \leq 3.50$ and the lower mass group spanning $0.95 \leq M/M_\odot \leq 1.50$. The overall period range of these progenitors span $-0.38 \leq \log_{10}(P/ \rm days) \leq 0.38$. For more massive progenitor systems, the fraction of systems appearing similar to 4U~1735-444 is higher, but the time they spend in this state ranges from $10^4$ years to $10^8$ years. The lower mass progenitors have a smaller range between $10^4$ and $10^6$ years.

The progenitor binaries of 2A~1822-371 are very similar to the progenitors of 4U~1735-444 and form the group that is adjoined to the progenitors of 4U~1735-444 at similar initial orbital periods. Their masses range  $0.95 \leq M/M_\odot \leq 2.90$ and the periods range  $-0.34 \leq \log_{10}(P/ \rm days) \leq 0.40$. For both 4U 1735-444 and 2A 1822-371 (but especially for the latter one), many progenitors may not exist in nature: the donors either experience RLOF at ZAMS or when the NS was formed (see Figure \ref{fig:short_period_split}). Observationally determining the effective temperature of the donors could greatly constrain the possible progenitors, as the higher mass donors group have higher temperatures at $T_{\rm eff} \sim 5000 K$ than the low mass donors with $T_{\rm eff} \sim 3900 K$.

A key attribute that can be seen in Figure \ref{fig:short_period_split} is the smooth transition from one observed LMXB's progenitors to another with overlap between systems. The higher mass progenitors with $M \approx 2.5M_\odot$ can, depending on the period of the binary system, produce all four of our short period LMXBs. At this initial donor mass 4U 1735-444 can be reproduced with an initial separation of $\log_{10}(P/ \rm days) = -0.36$. As the initial period increases to $-0.32 \leq \log_{10}(P/ \rm days) \leq -0.28$, we effectively reproduce 2A 1822-37.

\subsection{Medium Period}

\begin{figure*}
    \centering
    \includegraphics[width=\textwidth]{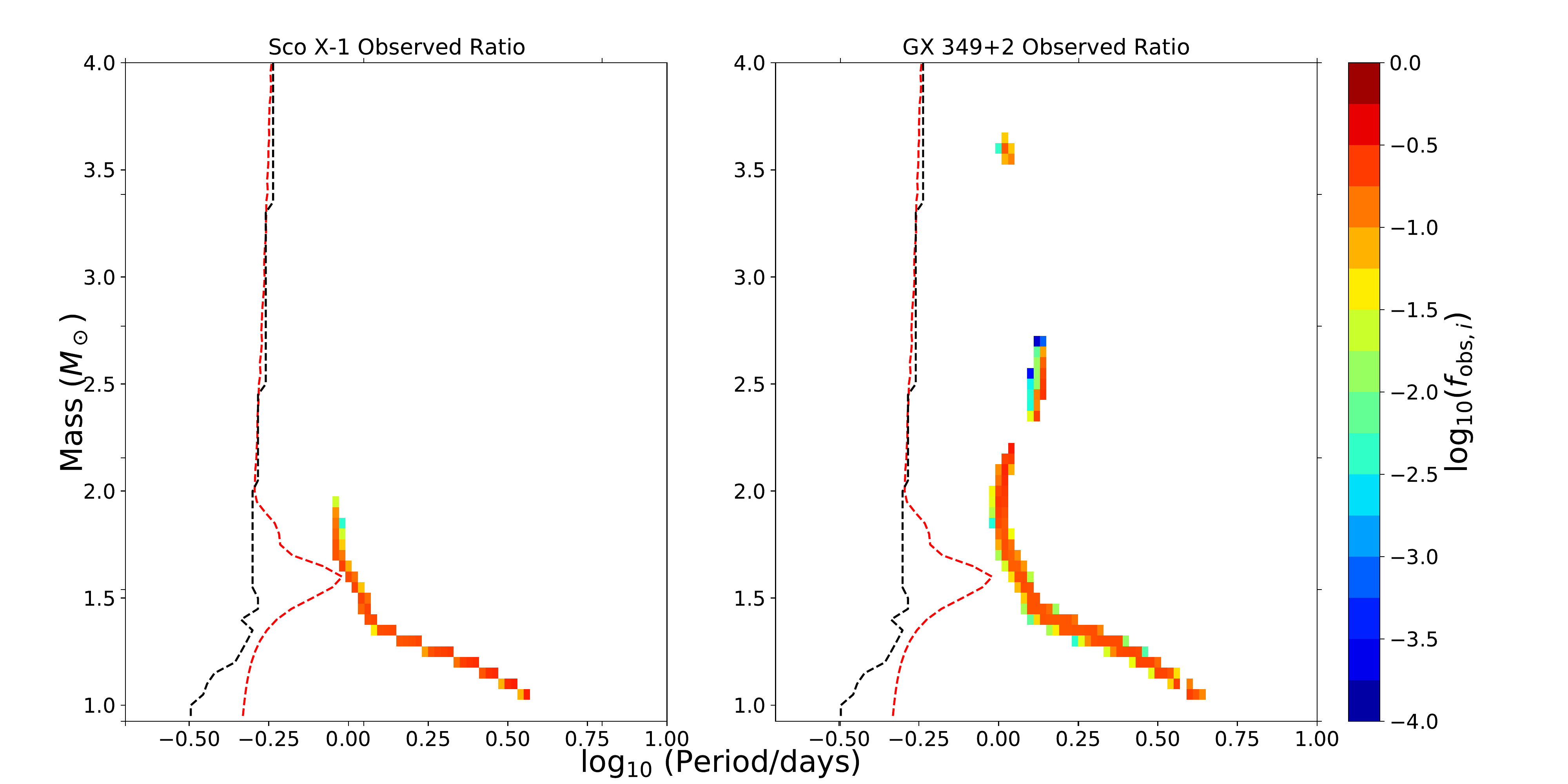}
    \caption{The ratio between observed and persistent times of the two medium period LMXBs. Similar to Figure \ref{fig:Progenitor_class}, the black and red dashed lines show the shortest period where an initial binary would start as a detached system and the shortest initial period which the detached binary can have when the progenitor is 10 million years old respectively.}
    \label{fig:med_period_split}
\end{figure*}

The progenitors of medium period LMXBs form a narrow strip in the initial period for each initial mass that spans over a large region of the periods (see Figure \ref{fig:Progenitor_Grid}).  Separating the progenitors of the two observed medium period LMXBs in Figure \ref{fig:med_period_split}, we see that Sco X-1 has the continuous range of possible progenitor masses, up to about 2 $M_\odot$, whereas GX 349+2 can have more massive progenitors, but not continuous in all masses above 2 $M_\odot$. The progenitors of Sco X-1 span nearly the same mass range as the low mass portion of GX 349+2 at slightly shorter, while almost adjacent,  periods. Specifically, the Sco X-1 progenitors have initial mass and period ranges of $1.05 \leq M/M_\odot \leq 1.95$ and $-0.04 \leq \log_{10}(P/ \rm days) \leq 0.56$. The low mass progenitors of GX 349+2 have similar masses, $1.05 \leq M/M_\odot \leq 2.20$. GX 349+2 has two gaps in the progenitor parameter space, one at $2.25 \leq M/M_\odot \leq 2.30$ and another between $2.75 \leq M/M_\odot \leq 3.55$. The lower mass gap is due to these systems not reaching the observed period, instead, these progenitors converge to an ultra-compact separation. The gap between $2.70 M_\odot$ and $3.55 M_\odot$ is due to the simulated binary not reaching sufficiently high mass transfer rates. 

We anticipate that Sco X-1 is more limited in our parameter space than GX 349+2 as we used the additional constraint on the donor's effective temperature. Without the effective temperature constraint, the mass range for Sco X-1 progenitors extends up to $2.55 M_\odot$, while with the effective temperature constraint, the upper mass limit remains at $2.0 M_\odot$. We discuss effective temperature more in section \ref{subsec:Teff}. If the effective temperature would be measured for the donor in GX 349+2, the mass of its progenitors could be constrained more, and either high mass or low mass formation channel could be ruled out.

The progenitors of GX 349+2 with the initial masses $2.35 \leq M/M_\odot \leq 2.70$ and the period of $\log_{10}(P/\rm days) = 0.12$ overlap with the progenitors of the UCXB 4U 1626-67, implying that GX 349+2 may evolve to a UCXB system. This small parameter space is the only initial parameter space with progenitors resulting in two LMXB systems with drastically different parameters during two different phases of mass transfer. 

\subsection{Long Period}

Cygnus X-2 is the only system in the long period category of our observed LMXBs. It is also one of the LMXBs with the observed effective temperature. This binary system stands out as there is a significant jump in the period from GX 349+2 to Cyg X-2 where the two LMXBs have observed periods of $\approx 22$ hours and $\approx 10$ days respectively. The progenitor parameter space area that can produce Cyg X-2 is comparable to the other observed systems (see Figure \ref{fig:Progenitor_Grid}). However, the maximum amount of time a simulated system spends in the appropriate bin (been similar to the observed LMXB) is much shorter. Specifically, the longest living binary spends less than $10^6$ years in the bin, but because the evolution is very rapid, the fractional time as defined in Equation \ref{eq:obs_frac_i} is large. The progenitor's initial space has less of a clear pattern, unlike the LMXBs we discussed previously. While the progenitors of Cyg X-2 almost form a strip, with the initial period increasing as the initial donor mass increases, there are gaps in the masses for plausible progenitors. The gaps in our progenitor space with initial masses below $M = 4.0M_\odot$ do not reach sufficiently high mass transfer rates whereas the systems with $M \geq 4.0M_\odot$ have mass transfer rates that exceed $10^{-7} M_\odot \mathrm{yr}^{-1}$ when at the appropriate mass ratio and period values. There is also an outlier at $M = 4.4M_\odot$ and $\log_{10}(P/{\rm days})=0.58 $. It is possible that the interruptions in the progenitor space is due to the quantization of our parameter space and additional fine tuning may be required in finding the viable progenitors in these gaps. It has been previously proposed that Cyg X-2 requires an intermediate-mass donor star \citep{Podsiadlowski2000}, and our results fully confirm that only this formation scenario can work. 

\subsection{Effective Temperature}
\label{subsec:Teff}

\begin{figure*}
    \centering
    \includegraphics[width=0.49\textwidth]{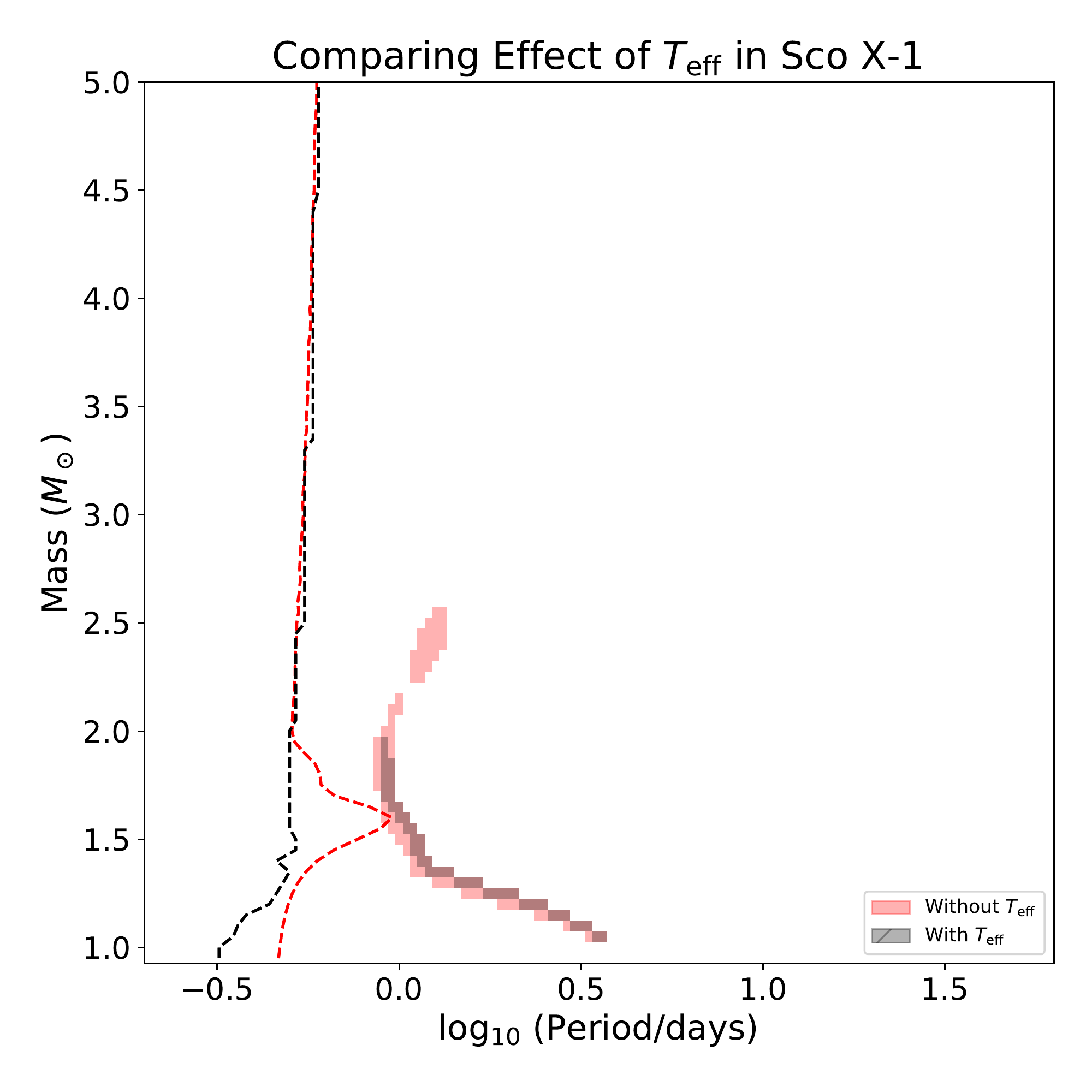}
    \includegraphics[width=0.49\textwidth]{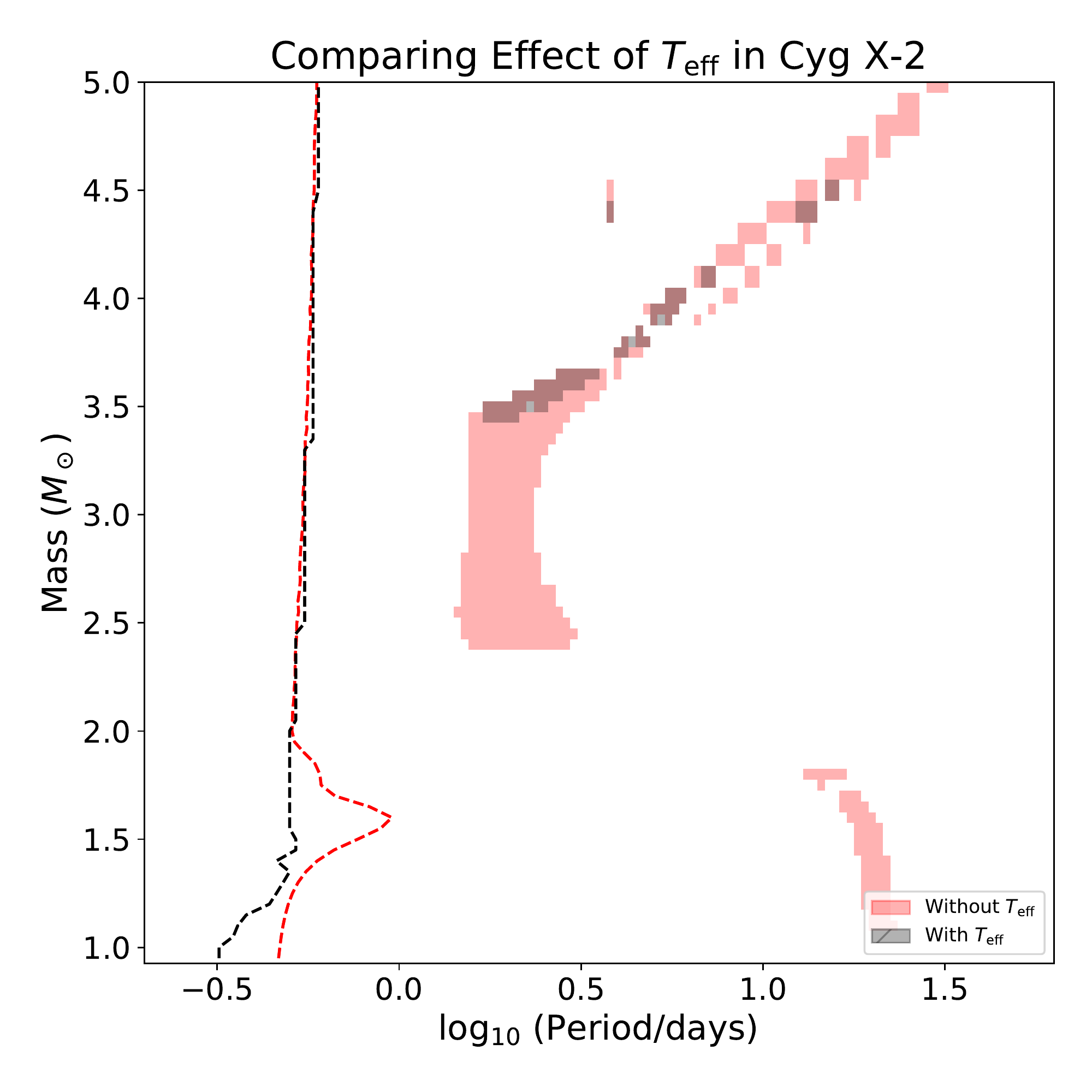}
    \caption{The progenitors of Sco X-1 and Cyg X-2 when the effective temperature is included in black and excluded in red. The black and red dashed lines are identical to those in Figure \ref{fig:Progenitor_class}.}
    \label{fig:Teff_Comp}
\end{figure*}

For Sco X-1 and Cyg X-2, we have an additional constant, the effective temperature, to limit the number of possible progenitors. Sco X-1 was found to have a donor star that has a spectral class later than K4 giving an approximate upper limit of $T_{\rm eff} \lesssim 4800$ K \citep{Sanchez2015}. Cyg X-2, on the other hand, ranges between an A5 to F2 spectral type, giving it a temperature range of $7000 - 8500$ K \citep{Cowley1979}. The inclusion of effective temperature as an additional constrain has significant effects on the possible progenitor systems, as can be seen in Figure \ref{fig:Teff_Comp}. 

\cite{Sanchez2015} places an upper limit on Sco X-1's effective temperature at $4800$ K. This temperature constraint limits the progenitor mass to $M_i \lesssim 2.0 M_\odot$ and narrows the width in initial periods at a given mass. The effective temperature plays a significantly larger role in limiting the progenitors of Cyg X-2 as seen in Figure \ref{fig:Teff_Comp}. Without the effective temperature, the number of viable progenitors for Cyg X-2 drastically increases and the viable initial masses can span the entire parameter space. Due to the defined range in $T_{\rm eff}$ for Cyg X-2, the excluded progenitors can have a wide range of possible values. The cluster of progenitors at $\log_{10}(P/ \rm\ days) \approx 1.3$ at lower masses results in LMXBs that appear similar to Cyg X-2 but the donor stars of these systems have significantly lower effective temperatures at $T_{\rm eff} \approx 4500 K$. The progenitors below $M_i \leq 3.5 M_\odot$ also do not reach sufficiently high temperatures with $T_{\rm eff}$ decreasing with progenitor mass. The progenitors with initial masses exceeding $M_i \geq 4.0 M_\odot$ on the other hand, consistently exceed the observed temperature with values of $T_{\rm eff} \geq 9000 ~\mathrm{K}$.

\subsection{Wider Bins}

\begin{table}
\footnotesize
\caption{\textbf{Extended MT for Non-UCXBs}}
\centering
\begin{tabular}{l | l}

System Name          & $\log_{10}(\dot{M}_a)$ \\
\hline

4U 1636-536          & [-9.50, -7.50] \\  
GX 9+9               & [-9.25, -7.50] \\  
4U 1735-444          & [-8.95, -7.50] \\  
2A 1822-371          & [-8.35, -7.00] \\  
Sco X-1              & [-8.45, -7.00] \\  
GX 349+2             & [-8.45, -7.00] \\  
Cyg X-2              & [-8.40, -6.90] \\  
\hline

\end{tabular}
\label{tab:mod_bins}
\begin{flushleft}
\textbf{Notes.} This table shows the extended mass transfer rate bins we use to determine the effect of our bin choices. The mass transfer rate is in units of $M_\odot\ \mathrm{yr}^{-1}$. The lower bound of our mass transfer rate is set to either $-9.5$ or 10\% the observed value, whichever is higher. The upper limit is then set to $-7.5$ or 0.5 dex above the observed value. The asymmetry in the limits is due to the Eddington limit which we do not expect most of these systems to exceed.
\end{flushleft}
\end{table}

\begin{figure}
    \centering
    \includegraphics[width=\columnwidth]{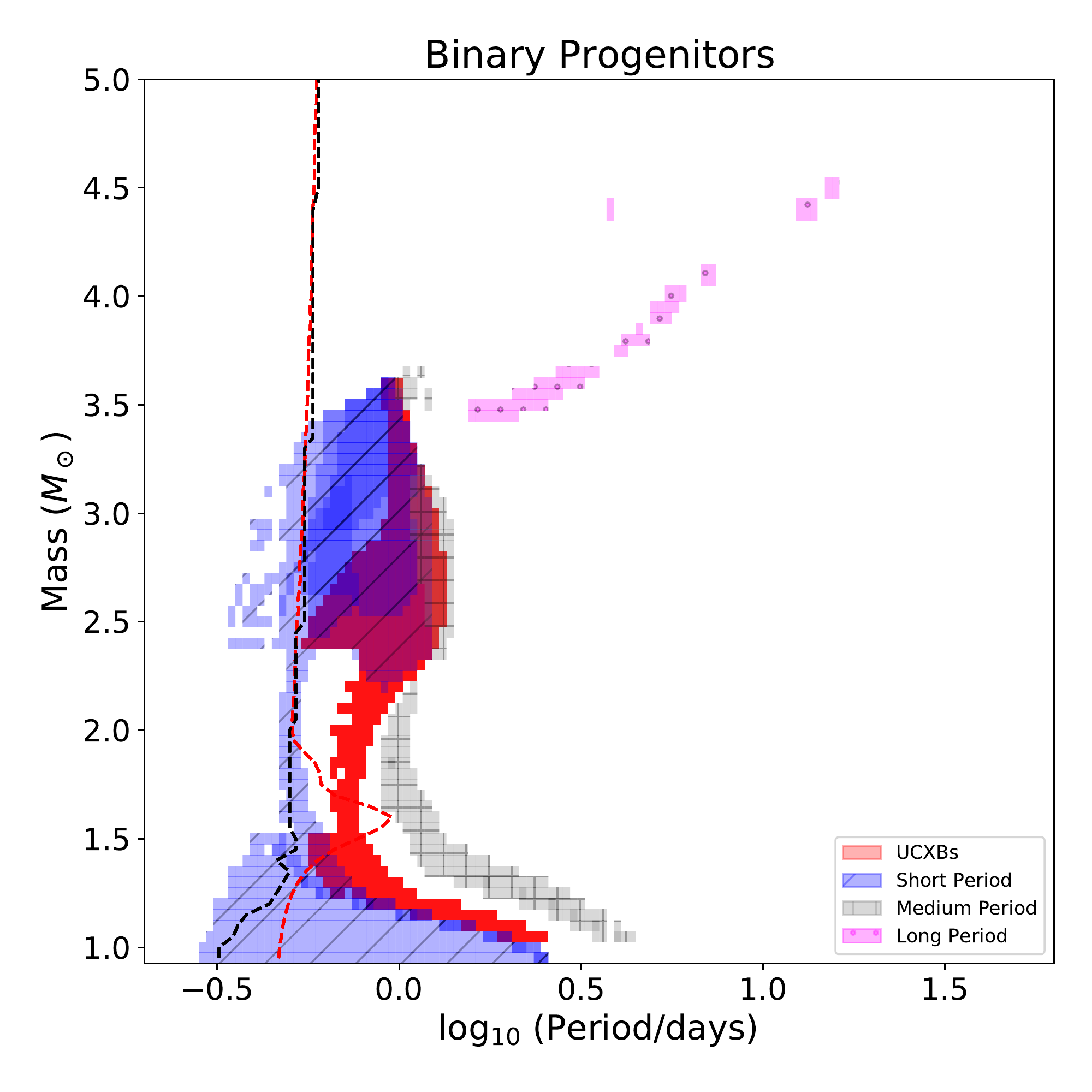}
    \caption{A similar figure to Figure \ref{fig:Progenitor_class} but using the wider bins for the observable quantities. The red and black dashed lines are the same as those described in Figure \ref{fig:Progenitor_class}. Again, many systems can share a common progenitor this results in some grid points being more saturated in colour if many observed LMXBs can be produced by that progenitor.}
    \label{fig:wide_progen_class}
\end{figure}

\begin{figure}
    \centering
    \includegraphics[width=\columnwidth]{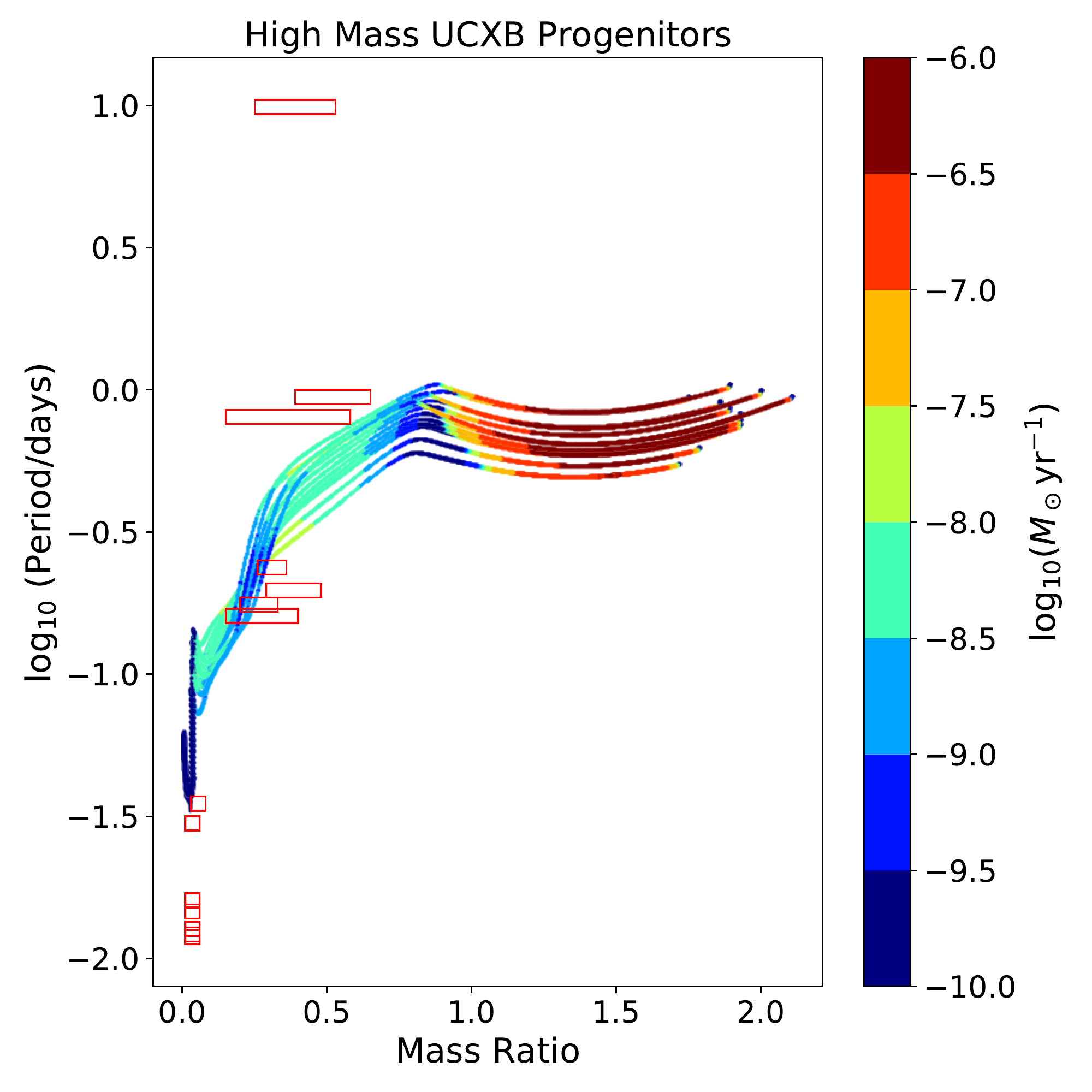}
    \caption{A sample of the high mass progenitors of UCXB systems found when using the wider bins. The red rectangles show the locations of the observed LMXBs using the wider bins. The colour bar shows the mass transferred $\dot{M}_{\rm tr}$}.
    \label{fig:high_mass_UCXB_progen}
\end{figure}

To determine the effect of the uncertainties in observed properties, we extend the size of our observed bins to encompass more significant errors. In particular, we focus on testing the period values for UCXBs and probe how varying the mass transfer rate changes our results. For UCXBs, we combine our bins to be $-2 \lesssim \log_{10}(P/\rm day) \lesssim -1.4$, $0.01 \lesssim q \lesssim 0.08$, and $-12 \lesssim \log_{10}(\dot{M}_a) \lesssim -7.5$. For our non-UCXB systems, we only vary the anticipated mass transfer rates. The extended values are given in Table \ref{tab:mod_bins}.

The progenitors that can create the LMXBs within the extended bins are shown in  Figure \ref{fig:wide_progen_class}. The most significant changes, as compared to our standard bins, can be seen in the progenitors of UCXBs and the short period systems. There is little change in the medium and no changes to the long period progenitors. In the case of UCXBs, previous studies and our narrower bins predict that UCXBs could only be produced by systems with initial periods near the bifurcation period \citep{Podsiadlowski2002}. Using the wider bins, the possible progenitors extend well beyond the area near the bifurcation period and have significant overlap with short period progenitors.

Examining the evolution of the models with high initial masses $M \geq 2.0 M_\odot$, we find that they can reach appropriately low mass ratios to match with observed UCXBs. However, the mass transfer rates from the progenitors found using the wider bins drop below $\log_{10}(\dot M_{\rm tr}) = -10$ when the simulated binary reaches ultra-compact periods and no longer satisfy the condition for a persistent LMXB using Equation \ref{eq:DIM}. These simulations instead reproduce transient UCXBs, which are outside the scope of this work. Additionally, these binaries can only shrink to periods of $\log_{10}(P/\rm days) -1.6$ prior to expanding to longer periods as seen in Figure \ref{fig:high_mass_UCXB_progen}.

\begin{figure}
    \centering
    \includegraphics[width=\columnwidth]{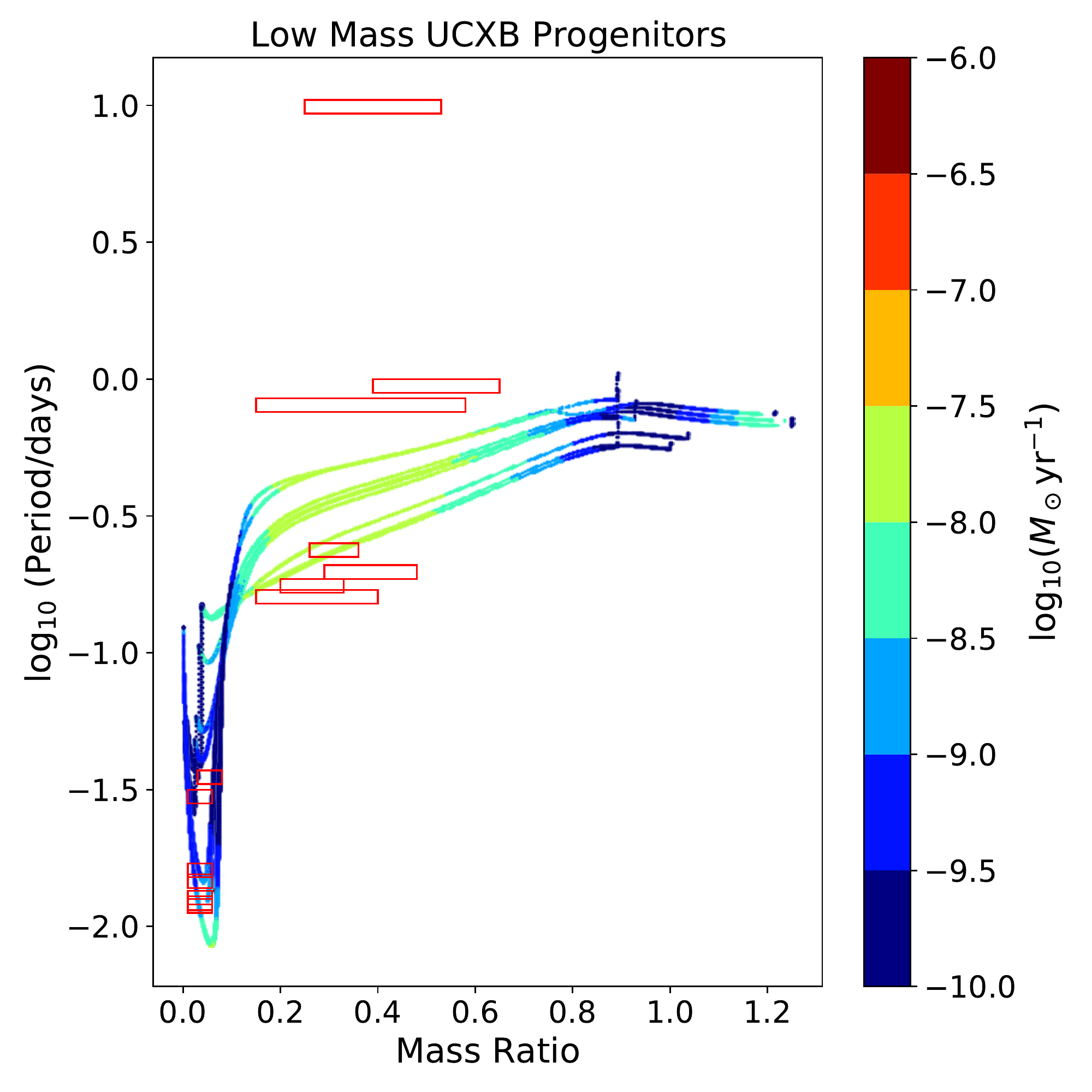}
    \caption{A sample of the low mass progenitors of UCXB systems found when using the wider bins. The red squares and colour bar are the same as in Figure \ref{fig:high_mass_UCXB_progen}.}
    \label{fig:low_mass_UCXB_progen}
\end{figure}

Similarly, the progenitors found using the wider bins with initial masses below $M < 2.0 M_\odot$ also reach appropriate low mass ratios and periods to match with UCXBs but similar to the high mass progenitors, do not reach sufficiently high mass transfer rates to satisfy Equation \ref{eq:DIM}. Again, the mass transfer rates drop below $\log_{10}(\dot M_{\rm tr}) = -10$ resulting in the simulated systems better matching with transient UCXBs. This can be seen in Figure \ref{fig:low_mass_UCXB_progen}. 

For the short period LMXBs, the extended MT bin results in a significant widening of the viable progenitors' initial periods. The new progenitors produce systems with periods and mass ratios similar to the observed systems, but the mass transfer rates exceed our standard bins at $\log_{10}(\dot{M}_\textrm{tr}) \approx -7.5$, extending the parameter space of 4U 1636-536, GX 9+9 and 4U 1735-444 as these systems have lower MT limits using the standard bins. The observed mass transfer rate appears to be the main constraint in reproducing short period LMXBs.

\section{Rate Estimates}
\label{sec:rates}

\begin{figure}
    \centering
    \includegraphics[width=\columnwidth]{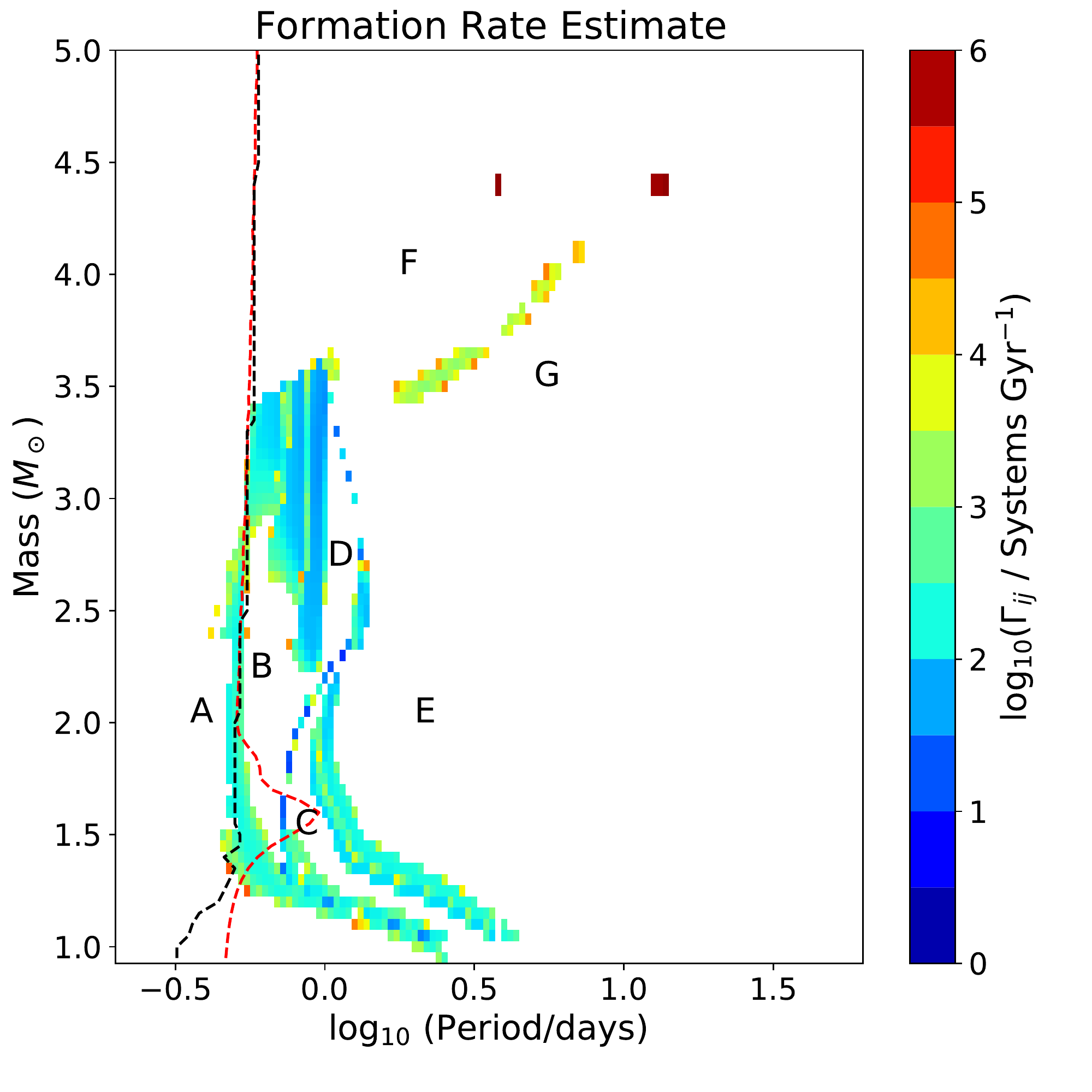}
    \caption{The minimum formation rate of the observed LMXBs, (see Equation \ref{eq:rate_calc}). The labelled sections in the plot denote specific regions of the parameter space where the simulations do not overlap with any observed LMXBs.}
    \label{fig:rate_density}
\end{figure}

\begin{table}
\centering
\begin{tabular}{l  cccc}

System Name          & $\log_{10}(P_i)$ & $M_{i} $  & $ \Gamma_{\rm min} $ & $ \Gamma_{\rm avg} $ \\
\hline
\multicolumn{5}{l}{UCXB} \\
4U 0513-40           & 0.08                         & 1.2        &   92               & 363        \\                 
2S 0918-549          & -0.12                        & 1.4        &   89               & 216        \\                
4U 1543-624          & 0.38                         & 1.05       &   69               & 432        \\                
4U 1850-087          & 0.16                         & 1.15       &   43               & 124        \\                
M15 X-2              & 0.36                         & 1.05       &   31               & 96         \\    
Combined 5 UCXBs     & 0.28							& 1.1		 &   123			  & 203		   \\	            

\\

4U 1626-67           & 0.06                         & 2.3        &   11               & 339        \\                
4U 1916-053          & 0.12                         & 2.6        &   17               & 760        \\                
\\
\multicolumn{5}{l}{Short period}  \\
4U 1636-536          & 0.0                          & 3.4        &   40               & 736        \\                 
GX 9+9               & -0.08                        & 3.3        &   60               & 530        \\                 
4U 1735-444          & -0.16                        & 3.45       &   93               & 2879       \\                 
2A 1822-371          & -0.32                        & 1.75       &   163              & 1317       \\                 
\\
\multicolumn{5}{l}{Medium period}  \\
Sco X-1              & -0.02                        & 1.65       &   106              & 317        \\                 
GX 349+2             & 0.04                         & 2.2        &   63               & 1039       \\                 
\\
\multicolumn{5}{l}{Long period}  \\
Cyg X-2              & 0.34                         & 3.5        &   1196             & 170631     \\                 
\\
\multicolumn{5}{l}{Combined} \\
With Cyg X-2         & --                           & --         &   2827             & 125913     \\
Without Cyg X-2      & --                           & --         &   1452             & 8766       \\
\hline
\end{tabular}
\label{table:rate_table}
\caption{The table shows the initial period and mass resulting in the minimum formation rate. The subscripts min and avg denote the minimum and average. The rate $\Gamma$ in units of systems per Gyr as calculated using equation \ref{eq:rate_calc}. The period is the initial orbital period of the progenitor binary in $\log_{10}(\rm days)$ and the mass, $M_i$, is the initial mass of the donor in $M_\odot$. The two combined rows at the bottom shows the average minimum and maximum calculated formation rates when we randomly select a progenitor for our observed LMXBs.}
\end{table}

In this Section, we will discuss the progenitor formation rates. Unfortunately, due to the small number of observations, only an estimate can be done. We consider each of the existing observations as a binomial process. In a simple case, if all the progenitors would be formed only in one ``bin'' $ij$, the total number of systems $N_{{\rm f},ij}$ that needs to be formed per unit of time $T$ for each given observed system $k$ is related to the expected number of observations $N^k_{\rm obs}$  as 
\begin{equation}
    N^k_{\rm obs} =  N_{{\rm f},ij}^k \frac{\tau_{ij}^k}{T}  
    =   \Gamma_{ij}^k \tau_{ij}^k \ .
\label{eq:rate_calc}
\end{equation}
Here $ij$ are the seed mass and period combination that produces the system $k$, $\tau_{ij}^k$ is the time spent in the observed bin of interest, and $\Gamma_{ij}^k=N_{\rm f,ij}^k/T$ is the formation rate of a given observed system $k$ from the bin $ij$.  In Figure \ref{fig:rate_density} we present $\Gamma_{ij}^k$ for each progenitor bin (calculations are made assuming that each $N^k_{\rm obs} =1$). If the bin produces two or more observed binaries, we provide the value of the formation rate that is associated with the largest $\tau_{ij}^k$. 

For each observed system $k$, with the adopted expected observation number $N^k_{\rm obs} = 1$, we can find the bin that produces the binary with the longest $\tau_{ij}^k$. 
Suppose all $k$-LMXB formations result from the progenitors coming from this bin only. In that case, the associated total formation rate will be the smallest of various realizations of LMXBs total formations rates, produced either individually or by any combination, of all the bins that produce the system $k$.  These formation rates, $\Gamma_{\rm min}^k$,  are provided in Table \ref{table:rate_table}. 

In the case when $m$ initial bins, with equal areas, are equally plausibly forming the observed system $k$, the binomial distribution can be written as

\begin{equation}
    N^k_{\rm obs} = N_{\rm f}^k \frac{1}{m} \sum_{m}  \frac{\tau_{ij}^k}{T} \ = \Gamma_{\rm av}^k \frac{1}{m} \sum_{m}  \tau_{ij}^k \ .
\end{equation}
$\Gamma_{ij}^k$ is the average formation rate of a given observed system $k$ from all possible bins $ij$. Our bins are not equally sized.
In the case when the progenitors are formed in $m$ initial bins, and the probability of creation of the progenitor-system $k$ is uniformly distributed over the initial parameter space in mass and period,

\begin{equation}
    N^k_{\rm obs}  = \Gamma_{\rm av}^k \frac{1}{\Delta A^k} \sum_{m}  \tau_{ij}^k dA_{ij}^k\ .
\end{equation}

Here  $\Delta A^{k}=\sum_m d A_{ij}^{k}$ is the total size of the progenitor region that produces the observed system, and $d A_{ij}^{k}$ is the size of the specific bin.
The average formation rates $\Gamma_{\rm av}^k$ that take into account the sizes of the specific bins are provided 
in Table \ref{table:rate_table}. 

We note of course that there is no guarantee that the seed binaries are uniformly distributed in mass and period space, after a supernova event, and after likely a common envelope event, which both strongly affected their previous evolution and birth masses and periods, and hence the average formation rates are just for a reference, to understand the possible spread in the expected formation rates, as compared to the minimum formation rates of the same systems. In general, both the minimum and average formations rates do not vary strongly between the different type of LMXBs, except for Cyg X-2. Below we analyze these numbers for individual observed systems, as well as for the population as a whole. 

\subsection{UCXBs}

We find that the average formation rates for UCXBs are the lowest among all type of LMXBs.
This is expected, as UCXBs with the observed parameters can live for a long time. For example, to produce systems similar to M15 X-2, an average formation rate of 96 systems per Gyr can explain its existence, the smallest average rate amongst our calculations. The progenitors that lead to the minimum formation rates are different for all the UCXBs with no clear trend in the initial properties that produce the minimum rate for the observed LMXBs.

The five UCXBs with the shortest period are very similar to each other. Anticipating this similarity, we consider a cuboid that encompasses all five systems and obtain the formation rates using for this cuboid that $N_{\rm obs} = 5$. The cuboid spans a period range of $-1.95 \leq \log_{10}(P \rm/ day) \leq -1.77$, mass ratio of $0.01 \leq q \leq 0.06$ and mass transfer rate of $-9.8 \leq \log_{10}(\dot{M}_{\rm acc} / M_\odot \rm yr^{-1}) \leq -8.2$. Due to the combined UCXB cuboid spanning a wider range in both period and mass transfer rate, the amount of time a simulation spends in this bin of interest is increased, resulting in a lower required formation rate. The minimum formation rate for the combined 5 systems is 123 systems per Gyr, and the average formation rate is 203 per Gyr. This is the rate required to form all 5 systems.

Among our observed UCXBs, a subset of them are detected within globular clusters. 4U 0513-40, 4U 1850-087, and M15 X-2 are observed within globular clusters and because of this, we note additional caveats in regards to the calculated formation rate. Unlike UCXBs in a low density environment, UCXBs in a high density environment like a globular cluster may be formed through dynamical encounters or physical collisions \citep{Verbunt1987, Ivanova2005, Ivanova2008}. \cite{Bildsten2004} calculated the expected UCXB formation rate through dynamical encounters as one per $2\times 10^6$ years per $10^7 M_\odot$ of globular clusters. This roughly translates to $\sim$ 500 UCXBs formed in $10^9$ years per $10^7 M_\odot$ if the formation rate remains largely constant. It is unclear if the formation rate within globular clusters can be compared to the values we have calculated in Table \ref{table:rate_table}.

Two UCXBs with the longest periods, 4U 1626-67 and 4U 1916-053, stand apart. To form them, a new evolutionary channel is required. It appeared that the minimum number of formations of progenitor systems required to explain their observations is the lowest on overall, only about a dozen in the Galaxy per Gyr. The average formation rate for 4U 1626-67 is similar to the other UCXBs, but 4U 1916-053 is the largest among UCXBs. For both of these higher period UCXBs, the lowest formation rate progenitors come from systems with initial periods $M_i \lesssim 1.4M_\odot$ with higher calculated rates coming from binaries where the initial mass $M_i \gtrsim 1.5M_\odot$.

\subsection{Short and Medium Period}

The formation rates for short and medium period LMXBs are comparable to that of UCXBs. The minimum formation rates are between 40 and 163 systems per Gyr. The average rates are a factor of 2-3 larger than for UCXBs ranging from $300 - 3000$ systems per Gyr. This suggests that whatever process in nature creates the progenitor binaries that later appear as short or medium LMXBs,  more of such progenitors might be necessary to be created than for UCXB progenitors, albeit not dramatically.  There is a possible trend in the value for the minimum formation rate for short period systems, such that the calculated rate is larger for longer period LMXBs. Medium period LMXBs do not follow that trend, however. The progenitors of one of the short period systems 2A 1822-371, is the only one that intersect with the black dashed line and neglecting the progenitors left of this dashed line results in a calculated formation rate of 164 systems per Gyr.

\subsection{Long Period}

The observed binary Cyg X-2 is the most anomalous system in the calculations. The progenitor mass, the progenitor period, and the required progenitor formation rates are all significantly higher than any other system. With the average progenitor rate of $\Gamma_{\rm min} \gtrsim 1.7 \times 10^5$ systems per Gyr, the formation rate needed to make Cyg X-2 is nearly two orders of magnitude larger than any other observed LMXB. The minimum formation rate is substantially more reasonable, $\Gamma_{\rm min} = 1196$ per Gyr. The high value of the minimum formation rate indicates that this system was unlikely formed by a random progenitor that can make Cyg X-2 (see Figure \ref{fig:Progenitor_Grid}), and has likely been formed by the progenitor resulting in the minimum rate, specifically, with the initial mass of $M = 3.5M_\odot$ and the initial orbital period of $\log_{10}(P/ \rm days) = 0.34$.
As is discussed below in section \ref{sec:unobserved}, systems formed in the parameter space near this point would create LMXBs that are overall similar, but a bit different, to Cyg X-2. It is also plausible that Cyg X-2 is just a random realization of such progenitors, and is not itself indicative of a specific intensive pre-LMXB formation channel. As shown in section \ref{subsec:Teff}, effective temperature plays a significant role in constraining the progenitors of high initial mass or high initial period binaries. From Figure \ref{fig:rate_density} we see the general trend that as the initial progenitor mass increases, the calculated rate also increases. 
Additionally, the progenitors of Cyg X-2 appear to have an optimal initial period for each given mass such that the formation rate is the lowest at that period and is increasing as the progenitor deviates more from the initial period.

\subsection{Combined Rates}

The summed minimum rate required to produce the observed sample of LMXBs is only about 3000 binaries per Gyr. In comparison, the average rate is not very useful as it is dominated by Cyg X-2 and requires significantly more systems per Gyr, most of which are donors with initial masses above $3.5 M_\odot$. The obtained formation rates encompass many orders of magnitude, making it difficult to make predictions of the actual formation rate.

Beyond UCXBs, it is unlikely that every observed LMXB is produced using the  progenitor that results in the lowest rate. To obtain a more realistic value, we randomly select one successful progenitor per observed LMXB to determine a combined rate. We do this random calculation 10000 times to find a distribution of possible values. The randomized calculated rates are presented in the bottom two rows of Table \ref{table:rate_table}. 
Unfortunately, randomly selecting a progenitor for each system (and hence using the whole range of possible rates) led to the predicted rate being dominated by the progenitor of Cygnus X-2. We reiterate that the  formation rates calculated for Cygnus X-2 are significantly larger than those required for other observed LMXBs with the minimum rate being an order of magnitude larger than any other calculated rate and the average being two orders of magnitude larger (see Table \ref{table:rate_table}). In both cases where we calculate a rate with and without Cygnus X-2, the minimum number of formed pre-LMXB systems is on a similar order of magnitude with a few thousand systems formed per Gyr but the average rate is significantly larger when Cyg X-2 is included. This minimum value is larger than the absolute minimum progenitor formation rate as it is calculated by randomly selecting progenitors. While the total number of systems that are formed with an NS can be much higher than our minimum number, as these systems do not need to be limited to the progenitor space, we stress that the minimum number that we have obtained (2827 with Cygnus X-2 and 1452 without Cygnus X-2) is the minimum number of successful NS binary formation events that can explain the observed population of LMXBs generated by randomly selecting progenitors. Our average values (125913 with Cygnus X-2 and 8766 without Cygnus X-2) are more likely to represent the formation rate that nature has to provide.

\section{Unobserved Systems}
\label{sec:unobserved}

\begin{table}
\centering
\begin{tabular}{l | l}

Region &  \\
\hline

A & Difficult to produce progenitor LMXBs.\\
B & Should produce LMXBs with \\
  & $-0.75 \lesssim \log_{10}(P/\mathrm{days}) \lesssim -0.25$, \\
  & $q \lesssim 1$, $\log_{10}(\dot{M} / M_\odot \rm \ yr^{-1}) \sim -7.5$  \\
C & Should produce LMXBs similar to Sco X-1. \\
D & Early portion of evolution short lived.  \\
  & End of evolution has no mass transfer. \\
  & Intermediate portion should be detectable \\
  & with $\log_{10}(\dot{M} / M_\odot \rm \ yr^{-1}) \sim -8.5$ \\
E & High mass ratio progenitors do not satisfy \\
  & condition from equation \ref{eq:DIM} for persistent \\
  & systems. \\
  & Low mass ratio systems should be observable. \\
F and G & High mass transfer rates result in short lived \\ 
  &binaries. \\

\end{tabular}
\label{table:unobs_table}
\caption{A summary of the different regions of our parameter space and a summary of how observable a system from this region is.}
\end{table}

Only part of the entire parameter space leads to the production of the observed LMXBs, see Figures \ref{fig:Progenitor_class} and \ref{fig:rate_density}. That space is not continuous, and there are gaps in the parameter space -- the binaries originating from those gaps did not produce a binary comparable to the sample of observed LMXBs we considered. In this Section, we investigate what happens to the binaries that started their evolution in the ``gaps''.
To take a closer look, we split the ``unsuccessful'' parameter space into seven distinct regions, see Figure \ref{fig:rate_density}.
Region ``A'' denotes the portion of parameter space where the progenitor binaries will start with very short periods, $P_i \lesssim 0.5$ days. 
The three regions between the UCXB, short period, and medium LMXBs are denoted by ``B'', ``C'', and ``D''.
Regions ``F'' and ``G''  represent regions with initial periods $P \lesssim 10$ days and $P \gtrsim 10$ days with high initial masses, $M_i \gtrsim 3.0M_\odot$ respectively. The region ``E'' is for seed systems with a long initial period and initial masses below $3.0M_\odot$. Below, we examine these regions to determine if these LMXBs have properties that result in systems that, in theory, could be observable.

\subsection{Region ``A'': Short Initial Periods}

\begin{figure}
    \centering
    \includegraphics[width=\columnwidth]{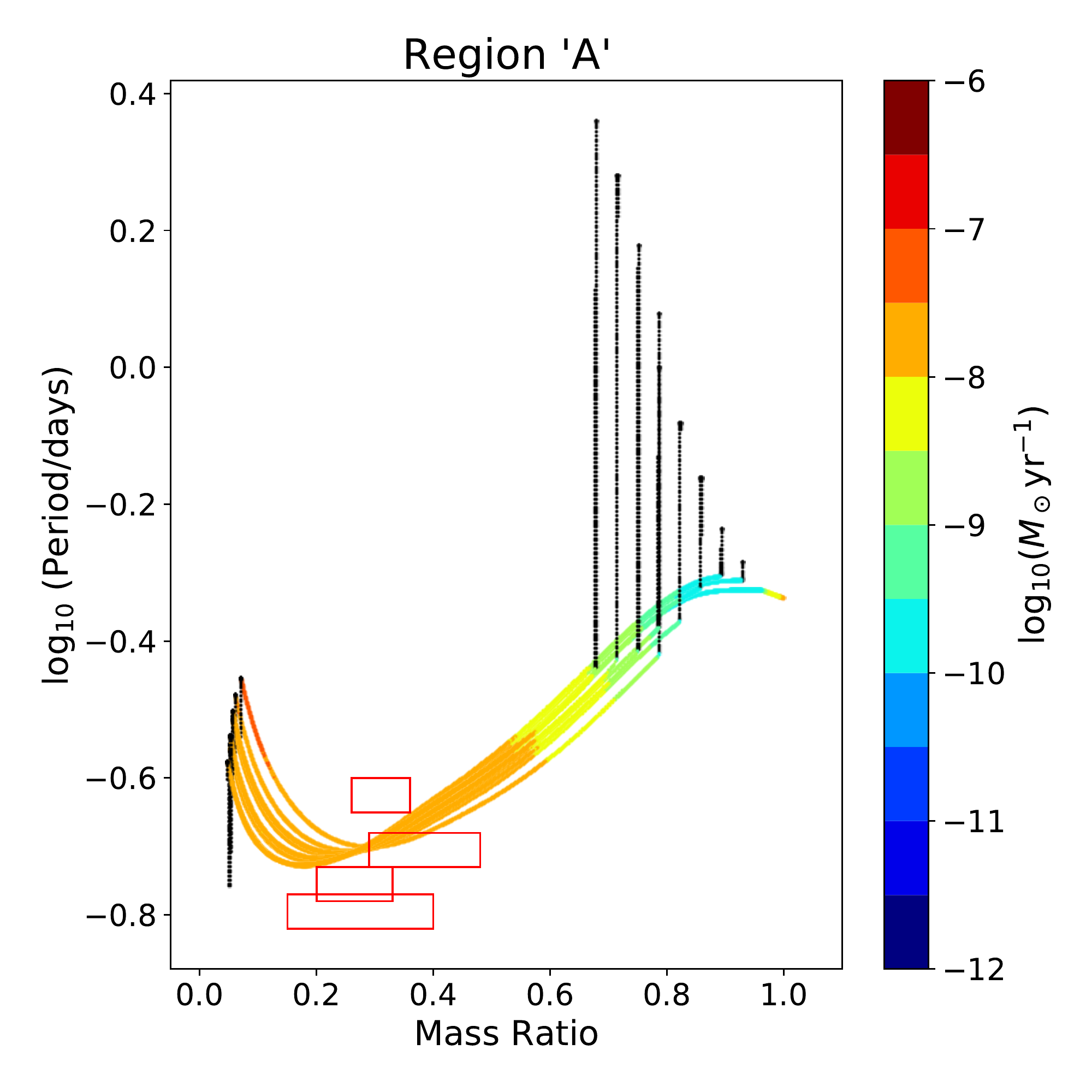}
    \caption{A subset of simulated systems with a short initial period, $P_i \lesssim 0.5$ days. The color of the lines indicates the mass transfer rate of a given simulation at that point. The red boxes show the mass ratio and period bins of the four observed short period LMXBs in Table \ref{table:combined_table}. The black points show portions of the evolution where the binary does not experience RLOF.}
    \label{fig:short_init_P}
\end{figure}

The short initial period region of our parameter space is primarily on the left of the black dashed line in Figures~\ref{fig:Progenitor_class} and \ref{fig:rate_density},  with only a small part of this region located to the right of the dashed line, for donors with $M_i \lesssim 1.3 M_\odot$. To reiterate, the black dashed line denotes the initial periods for which the simulated system with this donor mass is at its RLOF at the start of the simulation. Any system with a shorter initial period overfills the Roche lobe at the donor's ZAMS. For the binaries with $M_i \lesssim 1.3 M_\odot$, there is a portion of the parameter space where the system has to evolve from ZAMS to start the mass transfer, and we will focus on this parameter space.

In Figure \ref{fig:short_init_P}, we show the sample of the systems with the short initial periods and low masses.   
The simulated systems that have progenitors in this region of our parameter space result in binaries that only partially match with observed LMXBs. In general, the simulated systems have mass transfer rates that exceed the observed rates of 4U 1636-536, GX 9+9 and 4U 1735-444. 2A 1822-371 on the other hand, has sufficiently high mass transfer rates to match with the simulated systems, but the observed period of 2A 1822-371 exceeds the simulated periods. These short initial period, low initial mass progenitors produce LMXBs that have high mass transfer rates similar to 2A 1822-371 and similar mass ratios, but with shorter periods.
Additionally, these simulated systems spend between $10^7$ to $10^8$ years transferring mass. We can make an estimate of a formation rate required to produce one observable LMXB with $\log_{10}(P\rm /days) \sim -0.7$, $q \sim 0.3$ and $-7.5 \leq \log_{10}(M_\odot \rm~yr^{-1}) \leq -9$ using Equation \ref{eq:rate_calc}. With the number of possible progenitors in this region of our parameter space resulting in a system with similar properties, we calculate a progenitor formation rate of a few hundred per $10^{9}$ years.
Therefore, we may expect that there are observable binaries that have not been detected in this region of the parameter space. The limiting factor in our ability to detect these systems would be the formation rate of the progenitors. 

\subsection{Region ``B'': $M_i \sim 1.5M_\odot$, $P_i \sim 1$ day}

\begin{figure}
    \centering
    \includegraphics[width=\columnwidth]{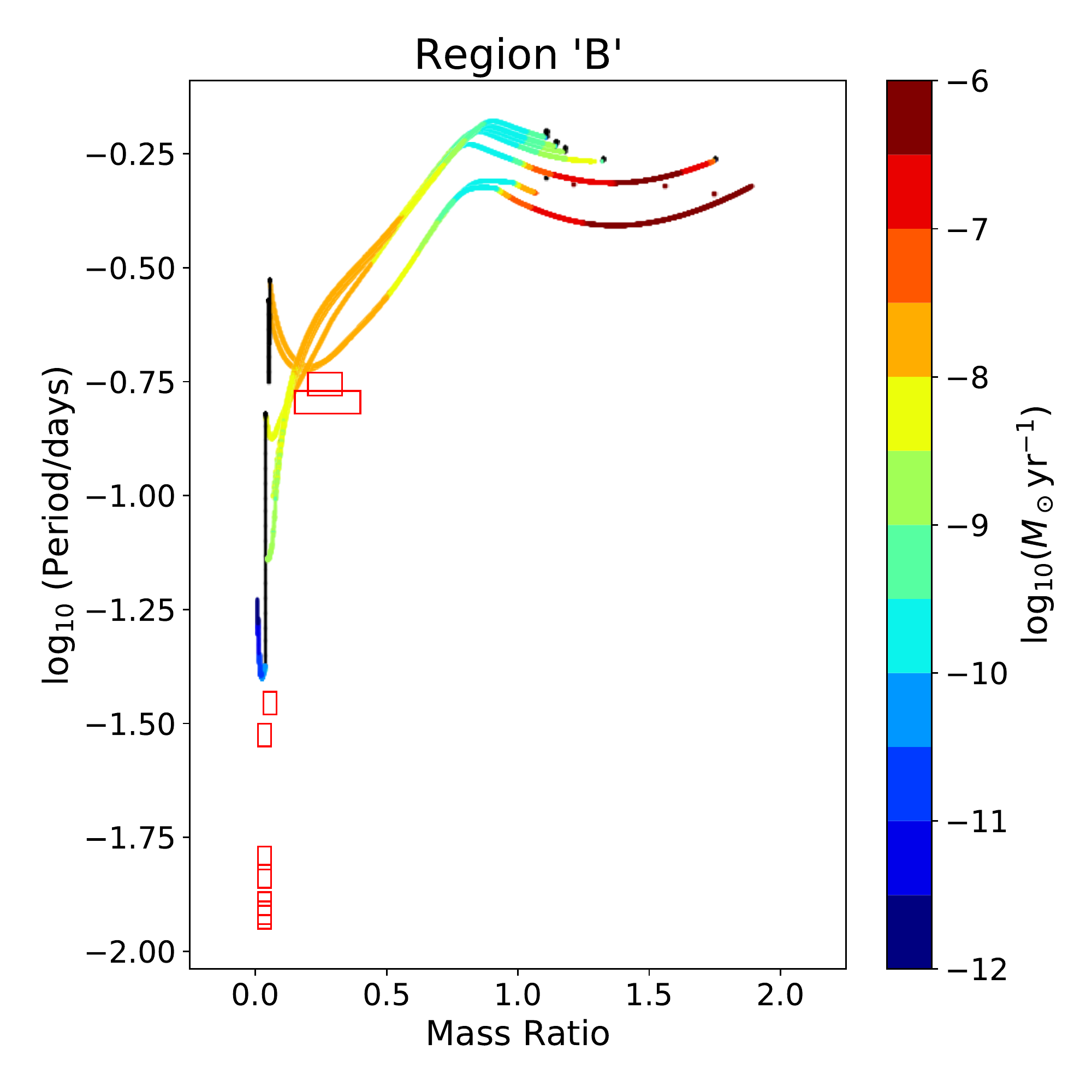}
    \caption{A subset of simulated systems from the region denoted as  ``B" in Figure \ref{fig:rate_density}.}
    \label{fig:B_super_short}
\end{figure}

This region of our parameter space is denoted by the letter  ``B" in Figure \ref{fig:rate_density} and lies between the progenitors of UCXBs and 2A 1822-371. The evolutionary tracks of the subset of the progenitor binaries can be seen in Figure \ref{fig:B_super_short}. The simulated systems initially experience high mass transfer rates, exceeding $\log_{10}(\dot{M} / M_\odot \rm \ yr^{-1}) \gtrsim -7$, with their orbital period remains almost unchanged and is around $\log_{10}(P / \rm days) \sim -0.3$ until the mass ratio flips. This initial phase of evolution is very short-lived and is therefore difficult to detect. Once the systems have reached a mass ratio of $\sim 1$, the evolution slows down, and the binaries are now long-lived. During this phase the mass transfer rate ranges between $-7.5 \leq \log_{10}(\dot{M} / (M_\odot \rm \ yr^{-1}) )\leq  -9$, while the orbital period of the binary decreases from $\sim 10 $ hours to $\sim 2$ hours. During this long period of evolution, the mass transfer rate consistently satisfies the condition for a persistent LMXB and should be observable. A binary with a period $\sim 8$ hours and mass ratio $\sim 0.5$ spend approximately $10^7$ years transferring mass $\log_{10}(\dot{M} / (M_\odot \rm \ yr^{-1}) ) \approx -8$. We predict that a minimum formation rate of a few hundred systems per Gyr in this region of the parameter space is necessary to produce an observable LMXB.

\subsection{Region ``C'': $M_i \sim 2.0M_\odot$, $P_i \lesssim 0.5$ days}

\begin{figure}
    \centering
    \includegraphics[width=\columnwidth]{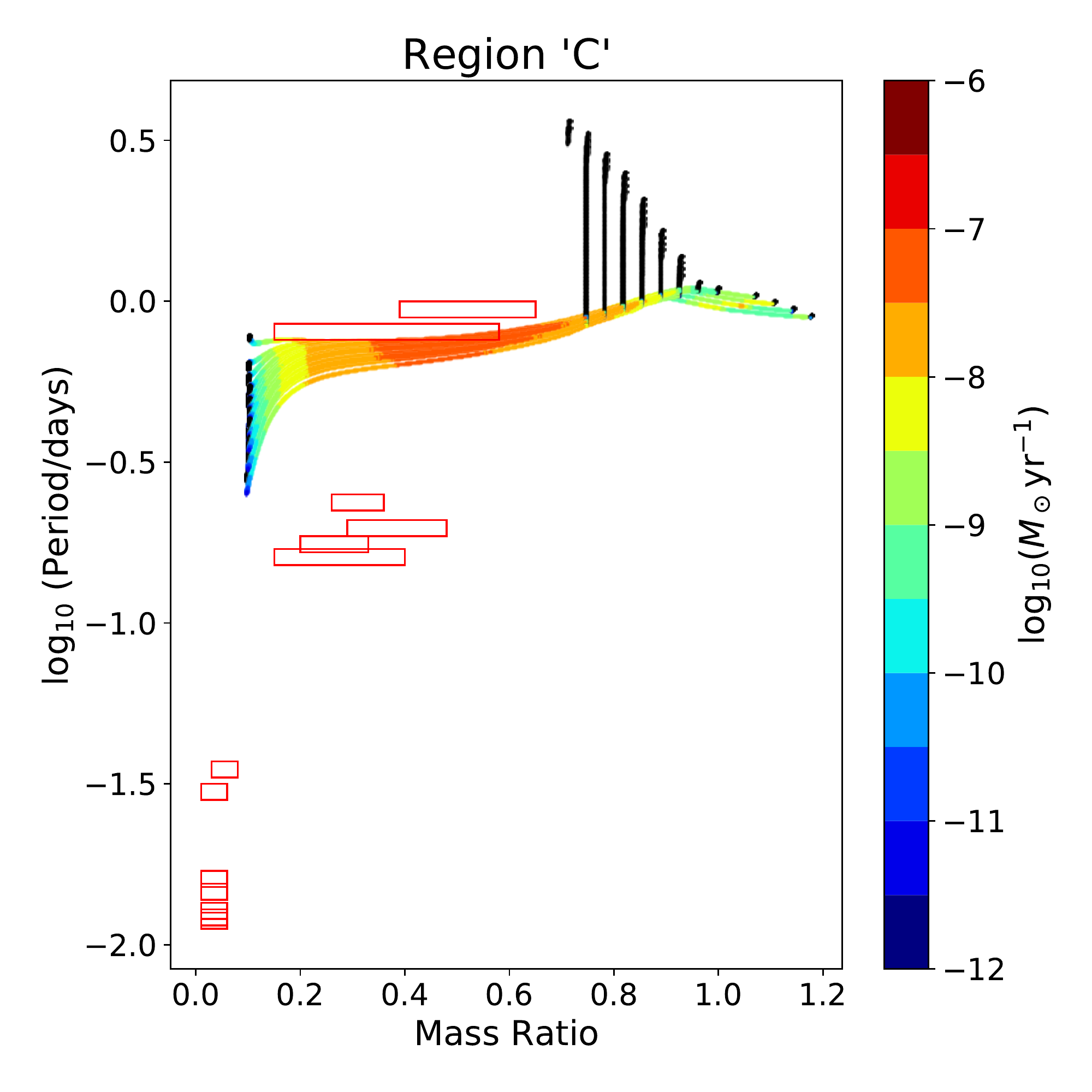}
    \caption{A subset of simulated systems from the region denoted as  ``C" in Figure \ref{fig:rate_density}. The red boxes show the bins for the UCXB and short period observed binaries.}
    \label{fig:C_super_UCXB}
\end{figure}

The simulations in region ``C'' of our parameter space have initial periods that exceed the bifurcation period but are not long enough to produce the medium period systems. The simulated systems in this region of our parameter space have very low mass transfer rates in the early stages and the end of their evolution, but significantly higher mass transfer rates in the middle of the evolution with the binary eventually detaching at the end of its evolution (see Figure \ref{fig:C_super_UCXB}). These simulated systems all converge in a similar region of the parameter space with orbital period, $P \sim 15~\rm hr$ with mass ratios in the range of $0.2 \lesssim q \lesssim 0.8$ and mass transfer rate $\log_{10}(\dot{M} / M_\odot \rm \ yr^{-1}) \sim -7.7$. These properties would result in an LMXB that is very similar to Sco X-1, but with a shorter period. The example systems that are shown in Figure \ref{fig:C_super_UCXB}, all spend on the order of $10^7 $ years in a persistent state with mass transfer rates exceeding $10^{-8} M_\odot \rm \ yr^{-1}$. Similar to Sco X-1, we would also predict a formation rate on the order of a few hundred systems per Gyr if the mass ratio bin remained wide. Further constraining the mass ratio to a range between $0.4 \lesssim q \lesssim 0.7$ where the mass transfer rate is highest at $\log_{10}(\dot{M} / M_\odot \rm \ yr^{-1}) \sim - 7.5$, the formation rate would need to increase to a few thousand systems per Gyr to produce and observed system.

\subsection{Region ``D'': $M_i \sim 3.0M_\odot$, $P_i \sim 1$ day}

\begin{figure}
    \centering
    \includegraphics[width=\columnwidth]{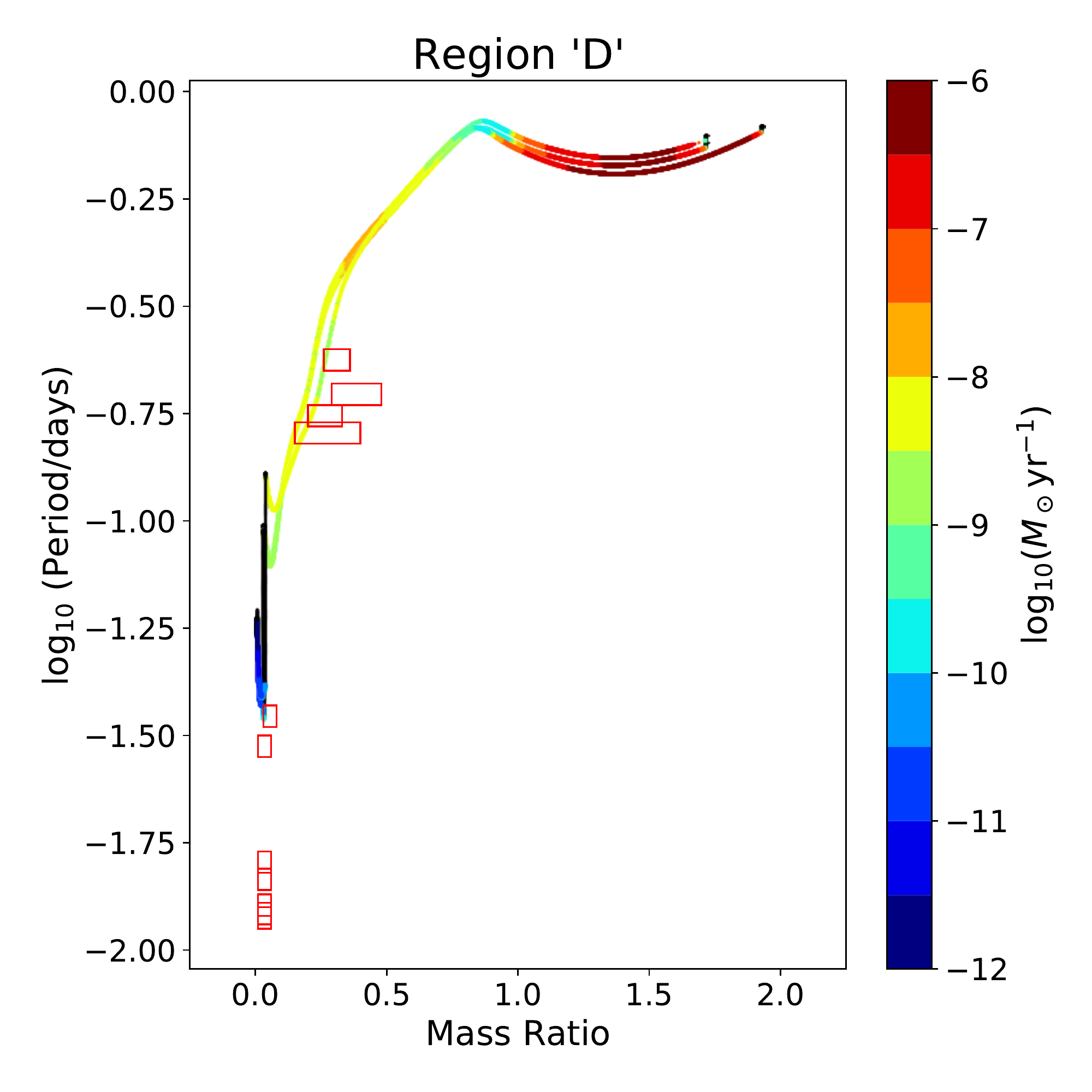}
    \caption{A subset of simulated systems from the region denoted as``D'' in Figure \ref{fig:rate_density}. The red boxes show the bins for the UCXB and short period observed binaries.}
    \label{fig:D_sub_UCXB}
\end{figure}

This region of our parameter space is denoted by the letter ``D'' in Figure \ref{fig:rate_density} and lies between the progenitors of UCXBs and GX 9+9. The evolutionary tracks of this subset of progenitors can be seen in Figure \ref{fig:D_sub_UCXB} and appears very similar to those in Figure \ref{fig:B_super_short} with these two regions forming a self-similar family of evolutionary tracks. These simulated systems also initially experience mass transfer rates that exceed $\log_{10}(\dot{M} / M_\odot \rm \ yr^{-1}) \gtrsim -7$ while the orbital period remains at around $P \approx 1$ day with their orbital period remains almost unchanged and is around $\log_{10}(P / \rm days) \sim -0.2$. This high mass transfer rate is short-lived and difficult to detect until the systems reach a mass ratio of $\sim 1$, where mass transfer slows down and the binaries are now long-lived.
Despite the mass transfer rate slowing down, they still remain large enough to be considered as persistent.
The mass transfer rates decrease to values between $-8.0 \lesssim \log_{10}(\dot{M} / M_\odot \rm \ yr^{-1}) \lesssim -9.0$ while the period decreases from $\sim 1 $ day to $\sim 2$ hours. Following the simulations in region ``B'', these progenitors also finish their evolution by detaching and becoming unobservable.

\subsection{Region ``E'': $P_i \gtrsim 1.5$ days, $M_i \lesssim 3.0M_\odot$}

\begin{figure}
    \centering
    \includegraphics[width=\columnwidth]{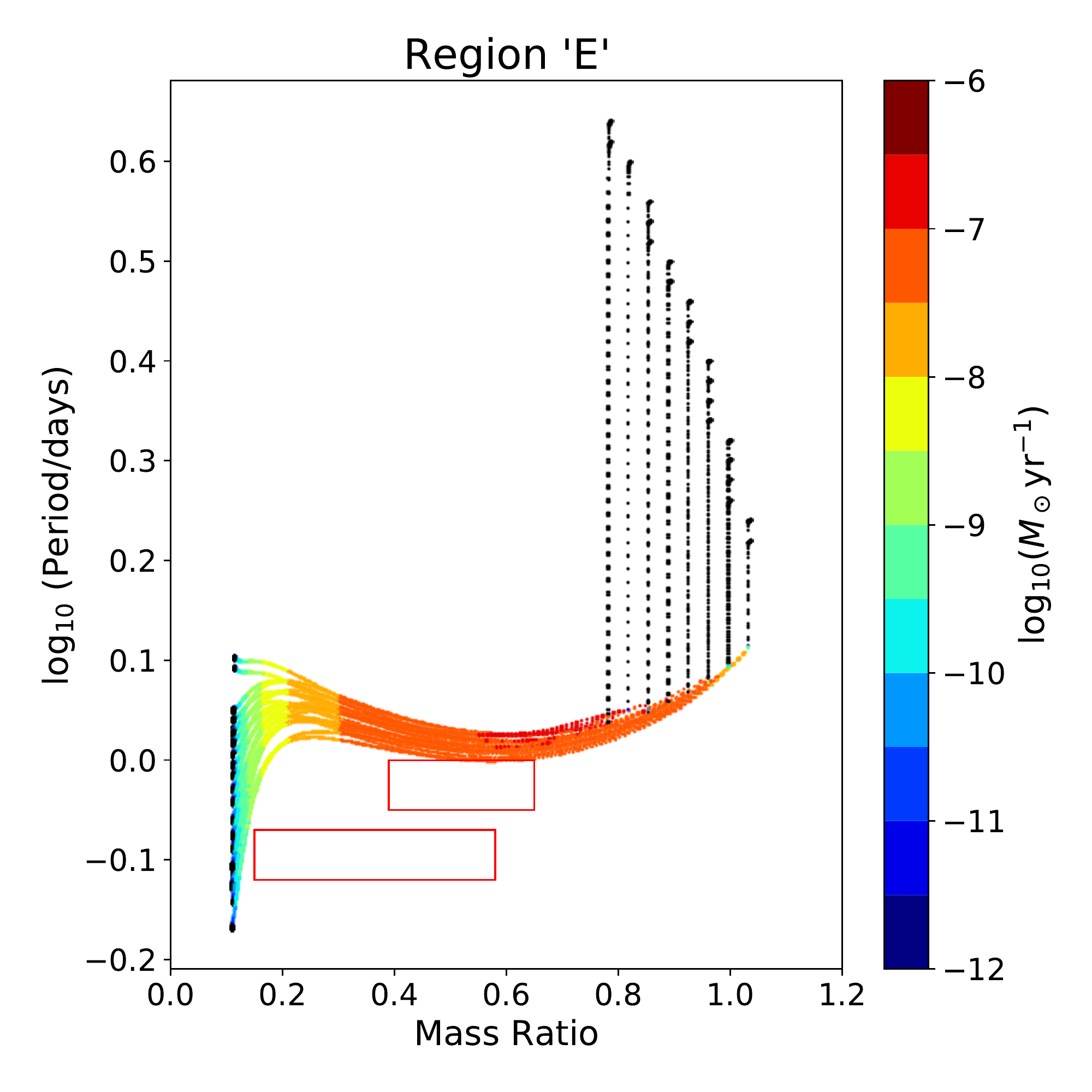}
    \caption{A subset of our parameter space such that the progenitor binaries have initial periods longer than 1.5 days and initial masses less than $3.0 M_\odot$ (region ``E''). The red boxes show the mass ratio and periods of the two observed medium period LMXBs in Table \ref{table:combined_table}.}
    \label{fig:long_init_P}
\end{figure}

The binaries with $P_i \gtrsim 1.5$ days and $M_i \lesssim 3.0 M_\odot$ are in region ``E'', see Figure \ref{fig:long_init_P}. Binaries with high initial masses and long initial periods are covered in region ``G''. The binaries in region ``E'' initiate RLOF when the periods shrink to $\sim 1.3$ days and during mass transfer the mass ratio decreases from 1 to $\sim 0.1$ prior to detaching. The orbital period remains well constrained during this time at $\sim 1$ day. The mass transfer rates of these binaries are above the critical value necessary to be deemed a persistent system in accordance with Equation \ref{eq:DIM}. The key difference between these simulated systems and the viable progenitors of Sco X-1 and GX 349+2 is that the simulated systems in region ``E'' have periods that are too large to match with our medium period systems. The simulated systems spend $\sim 5 \times 10^6$ years with $\log_{10}(P \rm / day) \sim 1$ day, $0.2 \leq q \leq 0.7$ and $-9 \lesssim \log_{10}(\dot{M} / M_\odot \rm \ yr^{-1}) \lesssim -7$ suggesting a very high minimum formation rate of a few thousand systems in this region per Gyr is necessary to produce an observable LMXB.

\subsection{Regions ``F'' and ``G'': High Initial Mass}

\begin{figure*}
    \centering
    \includegraphics[width=0.49\textwidth]{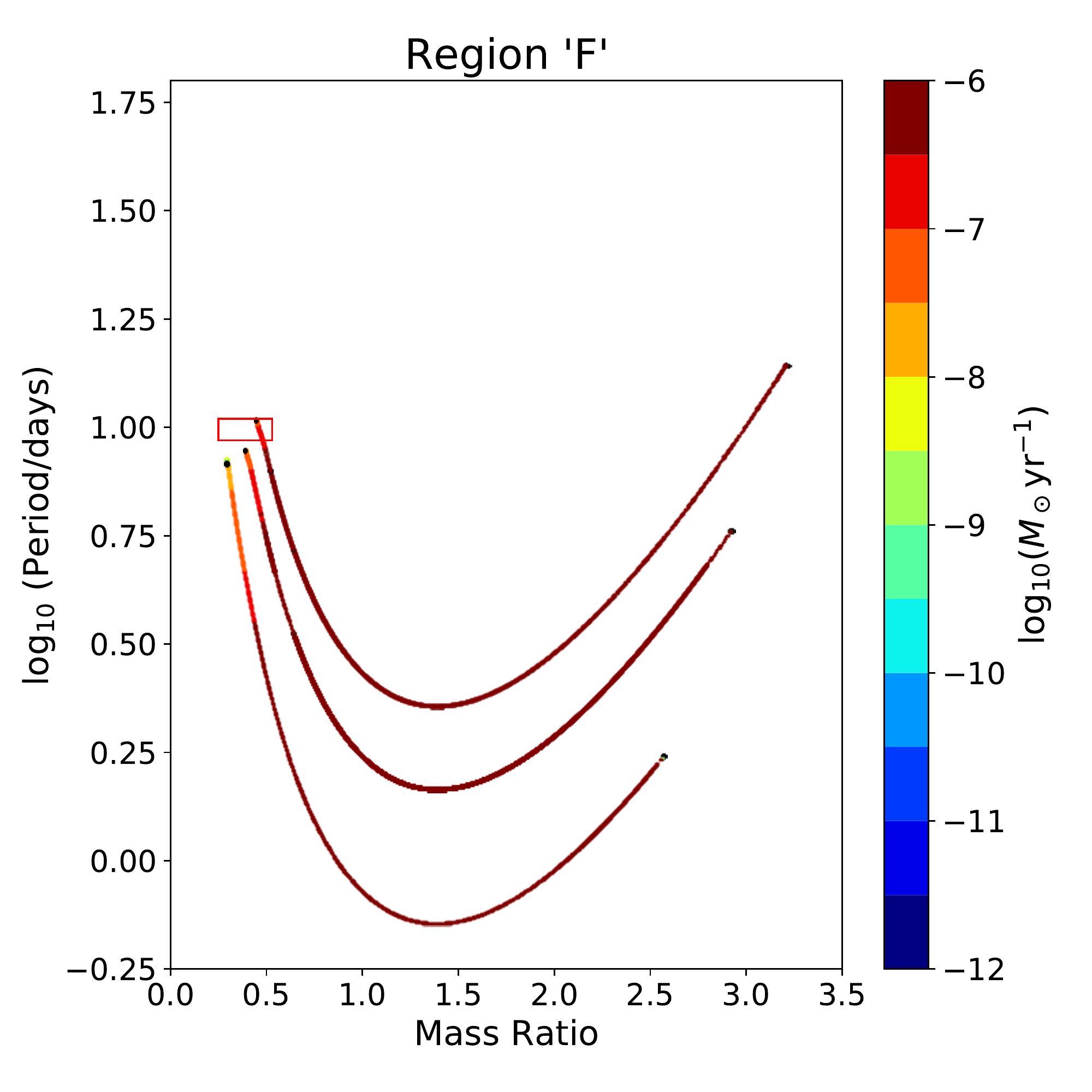}
    \includegraphics[width=0.49\textwidth]{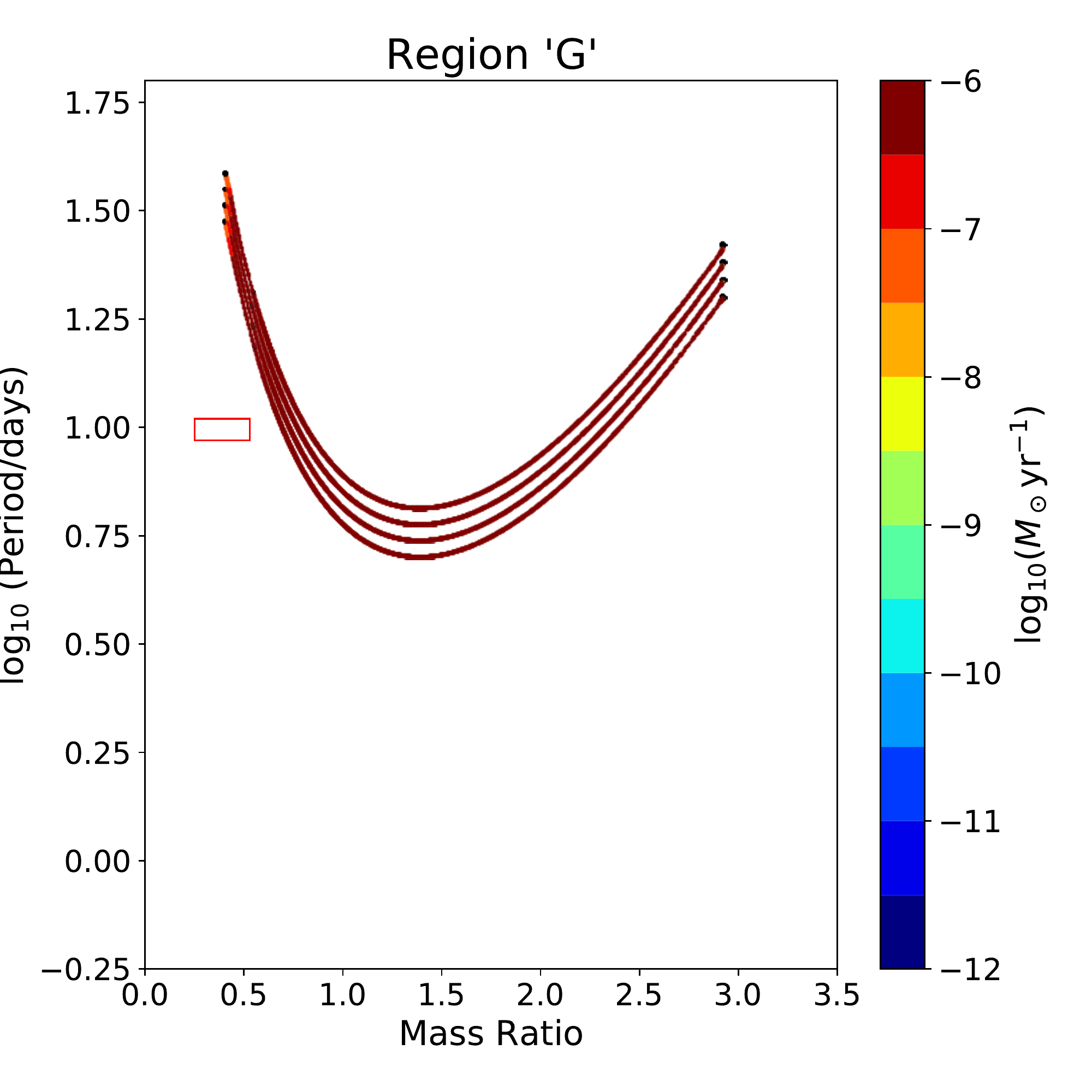}
    \caption{A subset of simulated systems with the initial donor masses exceeding at least $3.5M_\odot$, and more massive than Cygnus X-2 progenitors for each given initial orbital period.}
    \label{fig:high_init_M}
\end{figure*}

For binaries with high initial donor masses in region ``F'' and ``G'', these systems initially experience rapid mass transfer over $10^{-7} M_\odot \rm\ yr^{-1}$. With such significant mass transfer rates, the binary rapidly evolves to lower mass ratios, larger orbital periods, and smaller MT rates. Using Equation \ref{eq:DIM}, we find that once the orbital period is a few days or more, to be classified as a persistent binary, the required mass transfer rate must be $\log_{10}(\dot{M} / M_\odot \rm \ yr^{-1}) \gtrsim -7$.  However, a binary can not live long at such a high MT rate. These conflicting properties -- a short time while having a ``persistent'' MT rate and longer time while ``transient'' -- are the likely reasons why persistent binaries with large orbital periods are difficult to detect.  

\subsection{Uniform Initial Parameter Space Hypothesis}

As follows from the considerations for various ``unobserved'' regions above, there are regions that can also produce persistent systems. We do not have good guidance on the initial distribution of the seed binaries, which are themselves the results of the primordial binaries being evolved through a supernova, and likely a common envelope event. However, at the very least, we can test what happens if the seed binaries were formed uniformly in $\log_{10}P-M$ parameter space. We considered the random uniform birth of seed binaries in the $\log_{10}P-M$ parameter space. Further, we check if they can be observed (at the same fixed moment of time) based on their lifetimes as persistent systems. We made 1000 realizations that would consist of 14 persistent LMXBs. Out of our 1000 random realizations, over half don't produce any persistent systems that match with an observed LMXB. On average, $\lesssim$ 0.57 LMXB in such realizations would be consistent with the observed sample of the persistent LMXBs. The maximally consistent realization resulted in 3 LMXBs similar to the observed sample. Out of our 1000 realizations, 16 random progenitor sets result in 3 LMXBs similar to the observed sample. Averaging over these 16 sets, $2.3$ matched with short period systems, 0.3 matched with medium period and 0.4 matched with long systems. None of our random samples could reproduce an UCXB. The other persistent systems produced in these sets are most commonly found to originate from the regions `B', `D', `F' and `G'. Using this random sampling with 1000 realizations, we produce zero progenitor systems from the region `C' in our parameter space. The modelled but unobserved persistent LMXBs come from relatively high mass donors $M_i \gtrsim 2.5 M_\odot$. This suggests that uniform distribution of donors in mass in pre-LMXB binaries, after a supernova and presumably a common envelope event, is not likely in nature. We also note that the calculated number of LMXBs produced in a realization depends on the critical mass transfer rate in Equation \ref{eq:DIM}. By reducing this value by a factor of $\sim 2$ to $k=1.1\times 10^{15}$, this changes the number of LMXBs consistent with the observed sample to 0.56 which is not a significant change despite the change in critical mass transfer rate. Similarly, when increasing using the maximum value of $k=3.8\times10^{15}$ the value changes to 0.58.

\section{Conclusion}

Using the CARB MB from \cite{Van2019b} we obtained the possible progenitors of the observed persistent LMXBs with known orbital periods, mass ratio, and mass transfer rates. This was done by searching through the entire space of theoretically possible MT systems to infer possible initial conditions of the observed ones. Our results show that the viable progenitors of the observed LMXBs are located in a small part of the plausible parameter space, see Figure \ref{fig:rate_density}. The pattern of the progenitors' origins splits the persistent LMXBs that we analyze into distinct groups based on their currently observed periods.

Using these progenitors, we can calculate the minimum and the average formation rates, using the amount of time each progenitor spends appearing similar to an observed LMXB. The minimum formation rate is the value calculated using the simulated system that spends the largest amount of time matching an observed LMXB whereas the average formation rate uses random distribution among all plausible progenitors. The minimum formation rates for nearly all observed LMXBs are in the range of a few dozens to a hundred systems per Gyr in the Milky Way, while the average formation rates are in the range of a hundred to a few thousand systems per Gyr in the Milky Way.  Cyg X-2 is an outlier with a significantly higher range than all other systems. Without Cyg X-2, we find that the absolute minimum number of pre-LMXB binaries that needs to be formed per Gyr in the Milky Way to explain the observed sample of persistent LMXBs is about 1500 per Gyr. At the same time, a most expected number of seed binaries formation is about 9000 LMXBs per Gyr.

The key properties of the progenitors are as follows:

\begin{itemize}
    \item All UCXB systems have progenitors slightly below the bifurcation period. This period range is very narrow with the obtained minimum formation rates ranging between 11 to 92 systems per Gyr and the average formation rates ranging from 96 to 760 systems per Gyr.
    \item Short period LMXBs with periods on the order of a few hours have initial progenitor periods shorter than the bifurcation period with initial masses ranging from $\sim 1M_\odot$ to $\sim 3.5 M_\odot$. The short period LMXBs have the minimum formation rates similar to UCXBs, but our calculations predict the average formation rates much larger than for UCXBs. The average formation rates range between 530 and 2879 progenitor systems per Gyr. 
    \item Medium period LMXBs with periods ranging from tens of hours have initial periods slightly larger than the bifurcation period at lower masses, and share common progenitors with UCXBs at masses exceeding $\sim 2 M_\odot$. The minimum formation rates for medium period systems are very similar to UCXBs and short period systems, ranging from 63 to 106 systems per Gyr. The average formation rates are also similar, ranging from 317 to 1039 systems per Gyr.
    \item Cyg X-2, our only LMXB with an observed period on the order of tens of days, is an outlier. Unlike the other LMXBs we tested, progenitor space for Cyg X-2 lacks a clear structure. The rates necessary to reproduce Cyg X-2 are significantly higher than any other LMXB and dominates our calculations with the minimum formation rate ranging from $\sim 1200$ to $\sim 1.4 \times 10^6$, and the average rate of over $1.7 \times 10^5$ systems per Gyr. The high calculated formation rate indicates that Cyg X-2 is difficult to reproduce.
\end{itemize}

Of equal importance as our results showing the viable progenitors to observed LMXBs, is finding gaps in our progenitor parameter space where no observed systems have been detected. The main regions of interest are regions ``B", ``C", and ``D" in Figure \ref{fig:rate_density}. These progenitors' properties lie between the systems that produce observed LMXBs but have not resulted in any observable systems. This implies one of two possibilities, these progenitor properties are difficult to create in nature and no observed LMXBs exist, or these systems exist and simply have not been observed. Based on the properties of the systems from these progenitor regions, we have identified some observed LMXBs with partial observations that may match our simulations, examples of such systems are given in Table \ref{table:Possible_LMXB}.

Testing the scenario where seed binaries are uniformly formed in our parameter space, we check if this random distribution results in observable persistent LMXBs. Randomly generating 1000 realizations of progenitor systems that result in 14 persistent LMXBs, we find that over half don't produce any persistent systems that match with an observed LMXB and none of our 1000 realizations produce any UCXBs. Our random sampling also produces no progenitors from region `C' in our parameter space. The random sampling suggests that the unobserved persistent LMXBs would likely come from higher mass donors $M_i \gtrsim 2.5 M_\odot$ and so we can conclude that LMXBs seed binaries can not be formed uniformly in the donor mass. To explain the observed number of UCXBs, the initial distribution of periods of LMXB seed binaries can not be uniform either.

It is difficult to directly compare LMXB formation rates found in previous studies using population synthesis codes and rates found using our method. Our results determine a formation rate of LMXB progenitors that have already completed a common envelope event and supernova to avoid uncertainties with these events. Forward population synthesis codes on the other hand start with primordial binaries and must evolve through both and therefore include all uncertainties associated with common envelope events and supernova natal kicks. Due to huge sensitivity to these two events, it is not surprising to find overproduction of luminous X-ray binaries in the Galaxy up to a factor of a hundred, as compared to the observed LMXBs, with some variations of how a common envelope event is treated \citep[e.g.,][]{Pfahl2003}. \cite{VanHaaften2015} performed a population synthesis study focused on shorter period LMXBs with progenitor properties ranging between $0.7 \leq M_d/M_\odot \leq 1.5$ and $0.5 \leq P_i /\textrm{days} \leq 2.75$. Their simulations predict $\sim 40$ persistent LMXBs which is within a factor of two of the observed number of persistent LMXBs in the bulge. While the number of LMXBs systems per moment of time is comparable to ours, we cannot compare these results to our rates as \cite{VanHaaften2015} does not explicitly compare their simulations to any observed systems, or provide the formation rates of the pre-LMXB systems per Gyr. \cite{Shao2015} perform a comprehensive population synthesis study where the initial donor masses ranged from $0.3 \leq M_i / M_\odot \leq 6$ and period $0.2 \leq P_i / \textrm{days} \leq 1000$ to find an I/LMXB birthrate in the range of $9 \times 10^{-6} - 3.4 \times 10^{-5} \textrm{yr}^{-1}$. Their rate is compatible with the formation rate of binary millisecond pulsars that are predicted to descend from I/LMXBs. We find that overall our formation rates are comparable to theirs. However, we note that while their binary birthrate matches with both our rates and that expected for binary millisecond pulsars, the population of I/LMXBs in \cite{Shao2015} have mass transfer rates that are systematically lower than observations leading the authors to predict missing physics in the modelling of angular momentum loss in binaries.

\begin{table}
\centering
\footnotesize
\caption{\textbf{Possible LMXBs}}
\begin{tabular}{l | cccc}

System       & $\log_{10}(P)$ & $\log_{10}(\dot M_a)$ & Region\\
\hline

4U 1746-37   & -0.67        & -9.0           & D              \\
2A 0521-720  & -0.47        & -7.4           & B              \\
4U 1624-49   & -0.06        & -8.3           & C              \\

\hline
\end{tabular}
\label{table:Possible_LMXB}
\begin{flushleft}
\textbf{Notes.} Examples of possible LXMBs that would be produced in the gaps in our progenitor space. All three systems lack an observation for mass ratio and the mass transfer rates are approximate values. The mass transfer rate is in units of $\log_{10}(M_\odot \rm\ yr^{-1})$. The period is in units of $\log_{10}(\rm day)$.  References: A09 - \cite{Agrawal2009}, B04 - \cite{BalucinskaChurch2004}, B09 - \cite{Balman2009}, C12 - \cite{Coriat2012}, L05 - \cite{Lommen2005}, L07 - \cite{Liu2007}, S01 - \cite{Sidoli2001}, X09 - \cite{Xiang2009}
\end{flushleft}
\end{table}

Further observations of these systems to constrain the properties and, more importantly, determine an approximate value for mass ratio would confirm if the proposed systems in Table \ref{table:Possible_LMXB} match our simulations. Additionally, if we had more systems with well-constrained periods, mass transfer rates and mass ratios, we would be able to further compare our simulated results to observed LMXBs. 
As the number of observed LMXBs with defined mass ratios, periods and mass transfer rates increases, any progenitor in our parameter space that fails to match with an observed system implies that binaries with that specific initial mass-period configuration are not likely formed from the primordial binaries. With the upcoming Gaia DR3 containing binary systems, we expect the number of observed LMXBs we can compare to will greatly increase, allowing us to probe the progenitor parameter space more effectively.

\section*{acknowledgments}
We would like the thank the referee for helpful comments.
N.I. acknowledges support from CRC program and funding from NSERC Discovery under Grant No. NSERC RGPIN-2019-04277. 
K.V. acknowledges the helpful comments from Erik Rosolowsky.

\facility{ComputeCanada}

\software{\texttt{mesaSDK} \citep{Townsend2019},
  \texttt{ipython/jupyter} \citep{Perez2018},
  \texttt{matplotlib} \citep{Hunter2018},
  \texttt{NumPy} \citep{Vanderwalt2018},
  \texttt{MESA} \citep{Paxton2011, Paxton2013, Paxton2015, Paxton2018, Paxton2019}
}

\bibliography{bibliography}{}
\bibliographystyle{aasjournal}



\end{document}